\documentclass[journal=langd5]{achemso}

\usepackage{amsmath}

\usepackage{array}

\usepackage{booktabs}

\usepackage[utf8]{inputenc}

\usepackage{makecell}

\usepackage{multirow}

\usepackage{threeparttable}

\usepackage{upgreek}

\usepackage{xcolor}

\title{Influence of charges on the behavior of polyelectrolyte microgels confined to oil-water interfaces}

\author{Maximilian M. Schmidt}
\affiliation{Institute of Physical Chemistry, RWTH Aachen University, 52074 Aachen, Germany}

\author{Steffen Bochenek}
\affiliation{Institute of Physical Chemistry, RWTH Aachen University, 52074 Aachen, Germany}

\author{Alexey A. Gavrilov}
\affiliation{Physics Department, Lomonosov Moscow State University, Moscow 119991, Russian Federation}

\author{Igor I. Potemkin}
\affiliation{Physics Department, Lomonosov Moscow State University, Moscow 119991, Russian Federation}
\alsoaffiliation{DWI - Leibniz Institute for Interactive Materials, 52074 Aachen, Germany}
\alsoaffiliation{National Research South Ural State University, Chelyabinsk 454080, Russian Federation}

\author{Walter Richtering}
\affiliation{Institute of Physical Chemistry, RWTH Aachen University, 52074 Aachen, Germany}
\email{richtering@rwth-aachen.de}

\begin{document}

\clearpage

\begin{abstract}
The role of electrostatics on the interfacial properties of polyelectrolyte microgels has been discussed controversially in the literature. It is not yet clear if, or how, Coulomb interactions affect their behavior under interfacial confinement. In this work, we combine compression isotherms, atomic force microscopy imaging, and computer simulations to further investigate the behavior of pH-responsive microgels at oil-water interfaces. At low compression, charged microgels can be compressed more than uncharged microgels. The in-plane effective area of charged microgels is found to be smaller in comparison to uncharged ones. Thus, the compressibility is governed by in-plane interactions of the microgels with the interface. At high compression, however, charged microgels are less compressible than uncharged microgels. Microgel fractions located in the aqueous phase interact earlier for charged than for uncharged microgels because of their different swelling perpendicular to the interface. Therefore, the compressibility at high compression is controlled by out-of-plane interactions. In addition, the size of the investigated microgels plays a pivotal role. The charge-dependent difference in compressibility at low compression is only observed for small but not for large microgels, while the behavior at high compression does not depend on the size. Our results highlight the complex nature of soft polymer microgels as compared to rigid colloidal particles. We clearly demonstrate that electrostatic interactions affect the interfacial properties of polyelectrolyte microgels.
\end{abstract}

\section{Introduction}

The use of solid colloidal particles as emulsion stabilizers has been established well over a century ago. \cite{Pickering1907} While particle-stabilized emulsions show incredible long-term stability against coalescence, an increasing number of present-day applications only requires temporary stability of the emulsion followed by controlled inversion or breaking. \cite{Tang2015} Hence, additional demulsification steps and/or extensive chemical functionalization of the solid particles are often necessary.

A new class of smart stabilizing agents that possess an inherent responsive nature are stimuli-sensitive microgels. They are crosslinked, three-dimensional polymer networks that are swollen by the solvent. \cite{Karg2019, Plamper2017} Changes in their surrounding conditions, including temperature, \cite{Pelton2000, Senff2000} pH, \cite{Hoare2004, Saunders2008} or solvent quality, \cite{Kaneda2004, Kojima2013} can induce a reversible volume phase transition of the polymeric network and adjust its dimensions and properties. Microgels can spontaneously adsorb to fluid interfaces and lower the interfacial tension. \cite{Li2014, Monteux2010, Zhang1999} Moreover, microgel-stabilized emulsions can be broken on-demand just by addressing the responsiveness of the polymer networks themselves. \cite{Brugger2008, Destribats2011, Destribats2011_2, Ngai2005, Ngai2006, Richtering2012, Schmidt2011, Schmitt2013, Rey2020} 

Since microgels are soft colloidal objects, they behave fundamentally different under interfacial confinement as compared to solid particles of similar dimensions. While the latter conserve their shape and volume upon adsorption, microgels are strongly deformed and flattened once situated at the interface. The extent of deformation is determined by the balance between elasticity of the polymeric network and energy gain associated with the reduction of interfacial tension. \cite{Style2015, Mehrabian2016, Rumyantsev2016, Camerin2019} Consequently, the effective lateral diameter of a microgel at the interface is substantially larger than in solution. 

Furthermore, microgels obtained by precipitation polymerization typically display a non-uniform distribution of crosslinks within the network. \cite{Wu1994} Their morphology is characterized by a more crosslinked core that is enclosed with a lesser crosslinked corona of dangling polymer chains. \cite{Stieger2004} This core-corona structure in bulk leads to a "fried-egg"-like morphology at the interface, in which the deformed core is surrounded by a very thin polymer film. \cite{Geisel2012, Zielinska2016} The structural inhomogeneity of the polymeric network and, thus, the softness directly affects the two-dimensional phase behavior. \cite{Pinaud2014, Geisel2014, Geisel2015, Rey2016, Picard2017, Rey2017, Scheidegger2017, Scotti2019}

Although the influence of external factors is well studied for microgels in solution, less is known about their response when adsorbed to an interface. Strikingly, microgels whose size in bulk is temperature-dependent show similar lateral diameters at the interface below and above their volume phase transition temperature. \cite{Harrer2019, Bochenek2019} Polyelectrolyte microgels, which contain comonomers with ionizable functionalities that are sensitive to changes in pH, exhibit similar behavior at the interface. Although the presence of charged groups within the microgel induces strong swelling of the network in solution, \cite{Schroeder2015} at the interface, the in-plane dimensions and the protrusion height into the non-aqueous medium depend only weakly on the surrounding conditions. \cite{Geisel2012}

The role of electrostatics on the interfacial properties becomes even more surprising when considering monolayers of charged microgels, so that interactions between them become relevant. Solid colloidal particles that are charged experience long-range Coulomb repulsion upon compression of the monolayer in a Langmuir trough. \cite{Aveyard2000, Aveyard2002} However, for pH-responsive microgels adsorbed to fluid interfaces, the results are inconsistent. On one hand, Geisel \textit{et al.} showed that the compressibility depends on the charge density of the microgel. \cite{Geisel2014_2} Langmuir compression isotherms were shifted between the uncharged and charged state and, counter-intuitively, charged microgels appeared to be easier compressible. The authors proposed that electrostatics do not directly affect the compressibility of microgels, i.e., through charge repulsion, but instead indirectly through the different swelling properties of uncharged and charged microgels. On the other hand, Picard \textit{et al.} found no dependence of the compressibility and microstructure of the monolayer on pH or ionic strength. \cite{Picard2017} All compression isotherms superimposed independent on the subphase conditions. Both groups used microgels based on \textit{N}-isopropylacrylamide (NIPAM) copolymerized with either methacrylic acid (MAA) or acrylic acid (AA), but the ones employed by Geisel \textit{et al.} were considerably smaller when charged than the ones of Picard \textit{et al.} Nonetheless, the results from both studies are in clear contrast to the behavior observed for charged rigid particles and emphasize the importance of the softness of the investigated system. Obviously, a more detailed description of how electrostatic interactions affect the interfacial properties of microgels is necessary.

In this work, we further elucidate the influence of electrostatics on the behavior of microgels adsorbed to fluid interfaces and propose an explanation for the dissimilar observations made so far. We start by revisiting the role of charges on the interfacial properties of the same microgels that have been investigated by Geisel \textit{et al.} In addition to compression experiments, we extend the previous study by Langmuir-Blodgett-type deposition experiments at controlled surface pressures and visualization of the deposited monolayers complemented by quantitative image analysis. Following this, we perform the same type of experiments with polyelectrolyte microgels of dimensions similar to the ones in the work of Picard \textit{et al.} In the last part of our study, we synthesize highly charged microgels from NIPAM and itaconic acid (IA) and probe the influence of the amount of charged groups as well as ionic strength on the interfacial properties. Our results demonstrate that electrostatics do affect the behavior of polyelectrolyte microgels confined to fluid interfaces and that the size of the investigated system plays an important role. We show that, depending on the degree of compression, the compressibility of the microgels is governed either by in-plane or out-of-plane electrostatic interactions.

\section{Experimental section}

\subsection{Materials}

\textit{N}-isopropylacrylamide (NIPAM) and potassium peroxodisulfate (KPS) were acquired from Acros Organics. Dimethyl itaconate (DMI), \textit{N,N'}-bis(acryloyl)cystamine (BAC) and \textit{N,N'}-methylenebis(acrylamide) (BIS) were purchased from Sigma-Aldrich. Sodium dodecyl sulfate (SDS), potassium chloride (KCl), n-decane, aluminum oxide and isopropyl alcohol were obtained from Merck. Sodium hydroxide (NaOH) and hydrochloric acid (HCl) were bought from VWR Chemicals. If not stated otherwise, all chemicals procured from commercial sources were used without further purification. The water used for all purposes was of ultra-pure quality.

\subsection{Microgels}

Two differently-sized MAA-containing NIPAM-based microgels were already obtained in previous works. The smaller ones, in the following referred to as S-MAA microgels, were synthesized by standard precipitation polymerization with surfactant. The exact details can be found elsewhere. \cite{Geisel2012} These are the microgels Geisel \textit{et al.} previously investigated at the oil-water interface. \cite{Geisel2014_2} The larger ones, termed L-MAA microgels, were synthesized in a semi-batch process, in which a shell consisting of NIPAM and MAA was polymerized onto a pure NIPAM core microgel. \cite{Schmidt2011} Hence, these microgels possess a distinct core-shell architecture. Both, S-MAA and L-MAA microgels, contain similar weight fractions of MAA. 

Additionally, highly charged microgels, abbreviated HC-IA microgels, were obtained in a two-stepped process. First, uncharged precursor microgels were synthesized from NIPAM and DMI employing precipitation polymerization. Use of the non-ionizable dimethyl ester of itaconic acid allowed for incorporation of a large amount of comonomer into the microgels (here 25~mol\%), but without causing forming oligomers to be too hydrophilic to precipitate from the reaction solution. Therefore, a 500~mL three-necked round-bottom flask equipped with a rubber septum, reflux condenser and mechanical stirrer was charged with NIPAM (2.9507~g, 26.08~mmol, 69.5~mol\%), DMI (1.4837~g, 9.38~mmol, 25.0~mol\%), BAC (0.0490~g, 0.19~mmol, 0.5~mol\%), and BIS (0.2901~g, 1.88~mmol, 5.0~mol\%). The disulfide bond-containing monomer BAC was added to allow for post-modification, e.g., labeling with fluorescent dyes. The contents were dissolved in 290~mL of degassed water and heated to 80~$^\circ$C under constant stirring (300~rpm). Simultaneously, KPS (0.1624~g, 0.60~mmol) was dissolved in 10~mL of degassed water. After addition of SDS (0.1775~g, 0.62~mmol) to the reaction vessel, the polymerization was started through injection of the KPS solution and allowed to proceed for 6~h at 80~$^\circ$C under nitrogen atmosphere. It was cooled down to room temperature, filtered over glass wool, and purified by five centrifugation-redispersion cycles (average RCF: 70400, 1.5~h). The amount of DMI incorporated into the precursor microgels was determined using nuclear magnetic resonance spectroscopy. Signals were fitted with Gaussian functions to derive the relative integrals, \cite{Massiot2002} from which the DMI content was then calculated. More quantitative information are included in the supporting materials and Figure S1.

In a second step, the ester moieties within the uncharged precursor microgels were hydrolyzed to yield pH-responsiveness. Therefore, multiple stoichiometric amounts (based on the DMI content) of 1M NaOH were added to an aqueous solution of the purified precursor microgels. The solution was left to stir at room temperature for 48~h and was subsequently neutralized by addition of 1M HCl. The obtained highly charged microgels were purified by five centrifugation-redispersion cycles (average RCF: 70400, 1.5~h). The extend of hydrolysis was quantified using conductometric titration. Details can be found in the supporting information and Figure S2.

We paid special attention to the purification procedure of the HC-IA microgels to exclude any interfacial active impurities remaining from the synthesis. \cite{Pelton2004} The efficiency of purification by centrifugation-redispersion cycles is discussed in the supporting information and data of the surface activity of the supernatant are presented in Figure S3. Briefly, microgel samples can be considered free of contaminants after four cycles as the surface tension of the supernatant becomes indistinguishable from the one of pure water. \cite{Vargaftik1983}

\subsection{Dynamic light scattering experiments}

Dynamic light scattering was used to determine the hydrodynamic radius of the microgels in bulk aqueous solution. Measurements were performed with a light scattering setup consisting of a HeNe laser (633~nm, 35~mW, JDS Uniphase Corporation USA), goniometer (ALV/CGS-8F, ALV GmbH Germany), digital hardware correlator (ALV-5000, ALV GmbH Germany), two avalanche photo diodes (SPCM-CD2969, Perkin Elmer Inc. USA), light scattering electronics (ALV/LSE-5003, ALV GmbH Germany), an index-match bath filled with toluene and an external programmable thermostat (Julabo F32, Julabo GmbH Germany). Samples were measured at 20~$^\circ$C and the scattering angle was varied from 30$^\circ$ to 100$^\circ$ (5$^\circ$ increments). Data were collected in dual-cross mode with an acquisition time of 120~s per measurement. To avoid multiple scattering and to ensure free diffusion, all samples were highly diluted. Data were evaluated utilizing the second order cumulant fit method. Therefore, the first cumulant was plotted against the square of the scattering vector and fitted with linear regression. The slope of the fit corresponds to the diffusion coefficient from which the hydrodynamic radius was calculated using the Stokes-Einstein equation.

\subsection{Electrophoretic mobility measurements}

Measurements of the electrophoretic mobility were carried out on a Zetasizer (NanoZS, Malvern Instruments Ltd. England) at 20~$^\circ$C using disposable folded capillary cells (DTS 1070, Malvern Instruments Ltd. England). The data were analyzed employing the Smoluchowski approximation. All samples were measured at least five times and the values averaged.

\subsection{Langmuir trough measurements and deposition experiments}

All experiments were performed with a customized Langmuir trough setup (KNIC 220, KSV NIMA/Biolin Scientific Oy Finland) for the investigation of liquid-liquid interfaces. It allows Langmuir-Blodgett-type deposition experiments simultaneous to compression of the interface. The trough, as well as two movable barriers (KN 0045, KSV NIMA/Biolin Scientific Oy Finland), are made from Polyoxymethylene. Holes drilled into the upper part of the barriers allow free flow of the oil phase during compression of the interface. The available area with the barriers in place was $\approx$~402~cm\textsuperscript{2} before compression, while the minimum area after compression was fixed to $\approx$~45~cm\textsuperscript{2}. The surface pressure was monitored by a platinum Wilhelmy plate (KN 0002, KSV NIMA/Biolin Scientific Oy Finland) with a perimeter of 39.24~mm connected to an electronic film balance. The plate was placed parallel to the barriers. pH and ionic strength of the aqueous subphase were adjusted by HCl or NaOH and KCl. Decane was used as the oil phase because it is a non-solvent for NIPAM-based microgels, \cite{Schmidt2011, Geisel2015_2} and less volatile than other alkanes as, for example, hexane. It was filtered at least three times over basic aluminum oxide to remove any polar contaminants (the last filtration was done just before an experiment was started). Microgel stock solutions were prepared at 0.25~wt\% with pH and ionic strength adjusted to the respective subphase conditions. The stock solution was mixed with 10~vol\% of isopropyl alcohol before addition to the interface in order to facilitate spreading. The trough was connected to an external water bath (Julabo F12, Julabo GmbH Germany) and all measurements were conducted at 20~$^\circ$C.

A typical compression experiment proceeded as follows: After the trough and barriers were carefully cleaned, a fresh decane-water interface was created and checked for impurities. The microgel solution was added to the clean interface using a Hamilton syringe and left to equilibrate for 60~min. Then, compression of the interface was started by closing the barriers symmetrically with a speed of 10~mm~min\textsuperscript{-1}, decreasing the available area. The barrier movement was stopped automatically once the set minimum area was reached. The surface pressure was recorded throughout the whole experiment.

Deposition experiments followed a similar protocol. The barriers were closed with a speed of 10~mm~min\textsuperscript{-1}. Once a desired surface pressure was reached, it was maintained over the whole deposition experiment by allowing the barriers to automatically adjust their position. After a waiting period of 10~min, the cleaned substrate, a rectangular piece of ultra-flat silicon wafer (T-UFSIW-6, NanoAndMore GmbH Germany), was raised through the decane-water interface with a speed of 0.1~mm~min\textsuperscript{-1} and the monolayer was deposited onto an area covering $\approx$~1~cm\textsuperscript{2}. The slow deposition speed was chosen to avoid significant disturbance of the interface that could potentially affect the structure of the transferred monolayer. The substrate was mounted to the dipping arm at an angle of 25$^\circ$ relative to the interface to be close to the contact angle between decane and water on a silica substrate and, thus, reduce the formation of a meniscus. The arm itself is constructed in a way that it was always above the fluid interface to reduce distortion (only the attached substrate was immersed). All substrates were cleaned by sonication in an isopropyl alcohol bath and subsequent plasma treatment before their usage.

\subsection{Atomic force microscopy imaging}

Imaging of microgel monolayers in the dry-state at the substrate-air interface was performed using an atomic force microscope with closed-loop (Dimension Icon, Veeco Instruments Inc. USA). Data were acquired in tapping-mode using OTESPA probes (OPUS by MikroMasch, NanoAndMore GmbH Germany) with a tip radius of $<$~7~nm. The resonance frequency of the cantilever was 300~kHz and the force constant 26~N~m\textsuperscript{-1}. Images were captured at a size of 7.5~$\upmu$m $\times$ 7.5~$\upmu$m or 20~$\upmu$m $\times$ 20~$\upmu$m, depending on the microgel. The resolution was always 512~px $\times$ 512~px.

\subsection{Quantitative image analysis}

AFM images were processed either by the commercially available software \textit{NanoScope Analysis 1.9} (Bruker Corporation USA) or open-source analysis software \textit{Gwyddion 2.54}.\cite{Necas2012} Tilt was removed from the micrographs through flattening and the scale was manually adjusted by setting the respective minimum to zero height. The processed images were analyzed with a custom-written script for \textit{MATLAB} (MathWorks Inc. USA). \cite{Bochenek2019} First, the centers of microgels were determined utilizing the publicly available \textit{MATLAB}-version of the IDL particle tracking code by Crocker and Grier.\cite{Crocker1996} With the center positions of individual microgels, their nearest neighbor connections were computed by a Delaunay triangulation and Voronoi tessellation.

The number of microgels per unit area $N_{\text{Area}}$ was calculated by dividing the number of localized microgel centers $N_{\text{P}}$ by the area of the respective image $A_{\text{Image}}$. Both, $N_{\text{P}}$ and $A_{\text{Image}}$, were corrected to exclude microgels located at the edges of the image.

\begin{equation}
    N_{\text{Area}} = \frac{N_{\text{P}}}{A_{\text{Image}}}
\end{equation}

The radial distribution function $g(r)$ was determined to obtain information about the long-range ordering within the monolayer. It is defined as:

\begin{equation}
    g(r) = \frac{1}{N_{\text{P}}} \ \bigg \langle \sum_{i~\neq~k}^{N_{\text{P}}} \delta (r_i - r_k) \bigg \rangle
\end{equation}

\noindent where $r_i$ and $r_k$ are the positions of the $i$-th and $k$-th microgel, respectively, and $\langle \ \rangle$ indicates the radial average over all possible orientations.

Furthermore, the two-dimensional hexagonal order parameter $\Psi_6$ was calculated for information on the short-range ordering. It is given as:

\begin{equation}
    \Psi_6 = \bigg \langle \frac{1}{N_{\text{B}}} \ \bigg | \sum_{j~=~1}^{N_{\text{B}}} \exp (i 6 \theta_j) \bigg | \bigg \rangle
\end{equation}

\noindent where $N_{\text{B}}$ is the number of nearest neighbors and $\theta_j$ is the angle between a chosen reference axis and the vector from a microgel to its $j$-th nearest neighbor.

Lastly, the mean center-to-center distance $d_{\text{cc}}$ between microgels was extracted from the maximum of the first peak of $g(r)$. Therefore, the distance probability function obtained from the Delaunay triangulation was fitted with either a single or the sum of two Gaussian functions.

\begin{equation}
    d_{\text{cc}} = (r \ | \ \text{max} [(g(r)])
\end{equation}

To obtain better statistics for all parameters listed above, the data from multiple AFM images of the same deposition but taken at different positions on the substrate were combined and averaged or plotted and fitted accordingly.

\subsection{Computer simulations}

Dissipative particle dynamics (DPD) simulations were performed with the same model as reported earlier, \cite{Gavrilov2019} to shed some light on the behavior of charged microgels at a liquid-liquid interface. DPD is a version of coarse-grained molecular dynamics. \cite{Hoogerbrugge1992,Schlijper1995,Espanol1995,Groot1997} Macromolecules are considered in terms of the bead-and-spring model. Each coarse-grained bead usually represents a group of atoms. Interactions between beads are modeled by a bond stretching force (only for connected beads), a conservative force (repulsion), a dissipative force (friction), and a random force (heat generator).

The soft core repulsion between the $i$-th and $j$-th beads is defined as:

\begin{equation}
    F_{ij}^{\text{c}} =
    \begin{cases}
    a_{\upalpha\upbeta} \left( 1 - \frac{r_{ij}}{R_{\text{c}}} \right) \left( \frac{\text{\textbf{r}}_{ij}}{r_{ij}} \right), & r_{ij} \leq R_{\text{c}}\\
    0, & r_{ij} > R_{\text{c}} 
    \end{cases}
\end{equation}

\noindent where $\text{\textbf{r}}_{ij}$ is the vector between the $i$-th and $j$-th bead and $a_{\upalpha\upbeta}$ is the repulsion parameter if particle $i$ has the type $\upalpha$ and particle $j$ the type $\upbeta$. $R_{\text{c}}$ is the cutoff distance describing the size of each bead in real units. It is basically a free parameter that depends on the volume of real atoms each bead is representing. \cite{Groot1997} Usually, $R_{\text{c}}$ is taken as the length scale, i.e., $R_{\text{c}}$~=~1.

If two beads ($i$ and $j$) are connected by a bond, there is a simple spring force acting on them:

\begin{equation}
    F_{ij}^{\text{b}} = -K_{\text{b}} \left(r_{ij} - l \right) \frac{\text{\textbf{r}}_{ij}}{r_{ij}}
\end{equation}

\noindent where $K_{\text{b}}$ is the bond stiffness and $l$ is the equilibrium bond length. The following parameters were used: $a_{\upalpha\upalpha}$~=~80, $K_{\text{b}}$~=~50, $l$~=~0.5. The time step for integration of the equations of motion was equal to 0.01.

It was demonstrated that the Flory-Huggins interaction parameter $\chi$ has a linear dependence on the difference of the DPD repulsion parameters $\Delta a$ = $a_{\upalpha\upbeta}$~-~$a_{\upalpha\upalpha}$. \cite{Groot1997} The relation for $a_{\upalpha\upalpha}$~=~130 was obtained in a previous work: $\chi$~=~(0.273 $\pm$ 0.007) $\Delta a$. \cite{Gavrilov2016} A similar relation holds for $a_{\upalpha\upalpha}$~=~25; therefore, we can assume that the same is true for $a_{\upalpha\upalpha}$~=~80. Thus, in what follows, $\chi$~$\approx$~0.28~$\Delta a$ was used.

Electrostatic interactions were taken into account in an explicit way using the method described earlier. \cite{Gavrilov2016_2} Within this approach, the electrostatic force between two charged beads is calculated as:

\begin{equation}
    F_{ij}^{\text{e}} = \frac{q_i q_j}{4\pi\epsilon\epsilon_0}
    \begin{cases}
    \frac{\text{\textbf{r}}_{ij}}{r_{ij}^3} \sin^6 \left( \frac{2 \pi r_{ij}}{4D} \right), & r_{ij} < D\\
    \frac{\text{\textbf{r}}_{ij}}{r_{ij}^3}, & r_{ij} \geq D
    \end{cases}
\end{equation}

\noindent where $D$ is the damping distance. The method allows the use of the exact Coulomb potential at distances larger than $D$, but avoids charge overlapping due to the soft potential. The damping distance was set to $D$~=~0.65, which was shown to be a good choice for the number density of 3. \cite{Gavrilov2016_2}

A single microgel was placed on a liquid-liquid interface. The liquids were immiscible as the $\chi$-value between them was equal to 5.6 ($\Delta a$~=~20). The size of the simulation box was 56~$\times$~56~$\times$~72 (677376 beads with the number density~3). The microgel was spherical and had an ideal diamond-like subchain topology following previous works. \cite{Rumyantsev2016, Gumerov2016} The subchain length was equal to 8 and the total number of beads in the microgel was $N$~=~10641. Ionizable groups were considered explicitly. To that end, some fraction $f$ of the monomer units bore a charge $e$, and the same number of counterions with a charge -$e$ was added to the system to preserve the electroneutrality. The Bjerrum length was assumed to be $l_{\text{b}}$~=~0.7~nm (water at room temperature) and the parameterization $l_{\text{b}}$/$R_{\text{c}}$~=~1 was selected, which is similar to that used by Groot. \cite{Groot2003} The two liquids represent polar and non-polar solvents like water and oil and are referred to as W and O, correspondingly. To take into account (in an implicit way) that the liquids have different dielectric permittivity and the charges prefer to be in the more "polar" liquid (W), it was assumed that the charged beads (monomers and counterions) are repelled from the "non-polar" liquid (O). The $\chi$-value between the O liquid and the charged beads was chosen to be 7~($\Delta a$~=~25), ensuring that all the charges were always located in the W phase. The incompatibility between the non-charged monomer units of the microgel and the polar liquid was varied. All other $\chi$-parameters in the system were equal to 0.

\section{Results and discussion}

\subsection{Characterization of polyelectrolyte microgels in solution}

In this study, we employ three different polyelectrolyte microgel systems based on NIPAM copolymerized with carboxyl group-bearing comonomers. Two of the microgels were already obtained in former works. \cite{Geisel2012, Schmidt2011} Both contain MAA as ionizable comonomer, but differ significantly in size. The smaller S-MAA microgels are the ones that Geisel \textit{et al.} previously investigated at the oil-water interface, \cite{Geisel2014_2} while the larger L-MAA microgels are of similar dimensions compared to the microgels that Picard \textit{et al.} studied. \cite{Picard2017}

In addition, we used a two-stepped process to obtain highly charged HC-IA microgels that include a much larger fraction of ionizable groups than is usually feasible by direct precipitation polymerization. First, uncharged precursor microgels were polymerized from NIPAM and the dimethyl ester of IA. In a second step, the ester groups within the microgels were hydrolyzed to yield IA and, thus, ionizable functionalities. More details regarding the synthesis of the HC-IA microgels are provided in the experimental section.

Before conducting interfacial experiments, we determine the effects of electrostatics on the solution properties for all three microgels. Table \ref{Solution_properties} lists their hydrodynamic radius $R_{\text{h}}$, derived from \textit{dynamic light scattering}, and \textit{electrophoretic mobility} $\mu$ under the employed conditions.

\begin{table}[H]
    \footnotesize
    \centering
        \begin{tabular}{cp{0.0mm}cccccc}
        \toprule[1.0pt]
        & & \multicolumn{2}{c}{\textbf{S-MAA}} & \multicolumn{2}{c}{\textbf{L-MAA}} & \multicolumn{2}{c}{\textbf{HC-IA}}\\
        \addlinespace[2.5mm]
        & & $R_{\text{h}}$ & $\mu$ & $R_{\text{h}}$ & $\mu$ & $R_{\text{h}}$ & $\mu$\\
        & & {[nm]} & {[10\textsuperscript{-8}~m\textsuperscript{2}~V\textsuperscript{-1}s\textsuperscript{-1}]} & {[nm]} & {[10\textsuperscript{-8}~m\textsuperscript{2}~V\textsuperscript{-1}s\textsuperscript{-1}]} & {[nm]} & {[10\textsuperscript{-8}~m\textsuperscript{2}~V\textsuperscript{-1}s\textsuperscript{-1}]}\\
        \midrule[0.5pt]
        uncharged & & \multirow{3}{*}{146 $\pm$ 1} & \multirow{3}{*}{-0.10 $\pm$ 0.01} & \multirow{3}{*}{424 $\pm$ 4} & \multirow{3}{*}{-0.09 $\pm$ 0.01} & \multirow{3}{*}{-} & \multirow{3}{*}{-}\\
        (pH 3, & & & & & & &\\
        0.1 mM KCl) & & & & & & &\\
        \addlinespace[5.0mm]
        charged & & \multirow{3}{*}{-} & \multirow{3}{*}{-} & \multirow{3}{*}{-} & \multirow{3}{*}{-} & \multirow{3}{*}{214 $\pm$ 2} & \multirow{3}{*}{-1.49 $\pm$ 0.08}\\
        (pH 9, & & & & & & &\\
        100 mM KCl) & & & & & & &\\
        \addlinespace[5.0mm]
        charged & & \multirow{3}{*}{222 $\pm$ 2} & \multirow{3}{*}{-1.34 $\pm$ 0.03} & \multirow{3}{*}{577 $\pm$ 7} & \multirow{3}{*}{-1.99 $\pm$ 0.05} & \multirow{3}{*}{275 $\pm$ 5} & \multirow{3}{*}{-2.45 $\pm$ 0.04}\\
        (pH 9, & & & & & & &\\
        0.1 mM KCl) & & & & & & &\\
        \bottomrule[1.0pt]
        \end{tabular}
    \caption{Solution properties, hydrodynamic radius $R_{\text{h}}$ and electrophoretic mobility $\mu$, of the S-MAA, L-MAA and HC-IA microgels. All measurements were performed at 20~$^\circ$C.}
    \label{Solution_properties}
\end{table}

The carboxyl groups within the MAA-moieties of the S-MAA and L-MAA microgels are protonated at pH~3 since the value is below the p\textit{K}\textsubscript{a} of methacrylic acid (p\textit{K}\textsubscript{a}~=~4.68). \cite{Volgger1997}
The specific size of the microgels is determined only by the balance between network elasticity and solvent quality. The slightly negative electrophoretic mobility at pH~3 can be attributed to initiator fragments remaining in the polymeric network from the synthesis, as anionic initiators were used. 

Increasing the pH to 9 causes deprotonation of the carboxyl groups, and the networks expand due to electrostatic interactions and osmotic pressure of the counterions. The more negative electrophoretic mobility of the L-MAA microgels compared to the S-MAA microgels can be explained by their architecture. In the L-MAA microgels, the acid moieties are only incorporated into the shell, so charged groups are mostly located towards the microgels' surface.

The relative change in size between uncharged and charged state in three dimensions is smaller for the L-MAA microgels ($R_\text{h,charged}$/$R_\text{h,uncharged}$~=~1.36) as compared to the S-MAA microgels ($R_\text{h,charged}$/$R_\text{h,uncharged}$~=~1.52), while the absolute difference in size is almost twice as large for the L-MAA ($\Delta R_\text{h}$ = 153 nm) over the S-MAA microgels ($\Delta R_\text{h}$ = 76 nm). Nevertheless, comparing the size of the two types of microgels at the same charge state, the L-MAA microgels are more than 2.6 times bigger in bulk (uncharged: 2.9; charged: 2.6) than the S-MAA ones.

The HC-IA microgels are charged as well at high pH because pH~9 is above both the p\textit{K}\textsubscript{a} values of itaconic acid (p\textit{K}\textsubscript{a,1}~=~3.85 and p\textit{K}\textsubscript{a,2}~=~5.45). \cite{Volgger1997}
They show the highest negative electrophoretic mobility at pH~9 and 0.1~mM~KCl due to the large amount of incorporated acid group-bearing comonomer. At pH~9 and an increased salt concentration of 100~mM~KCl, the effect of charge screening is visible. Under these conditions, the Debye length is significantly shorter and, thus, electrostatic repulsion of charged groups within the microgels is less dominant. The overall size of the microgels is smaller, even though the pH is maintained.

All three microgels behave as expected. Once charges are present inside the networks, they are more swollen compared to their uncharged state. Increasing the ionic strength leads to charge screening and a smaller overall size as compared to the non-screened state.

\subsection{Investigation of polyelectrolyte microgels at oil-water interfaces}

\subsubsection{Compression isotherms}

\begin{figure}[H]
    \centering
    \includegraphics[scale=0.35]{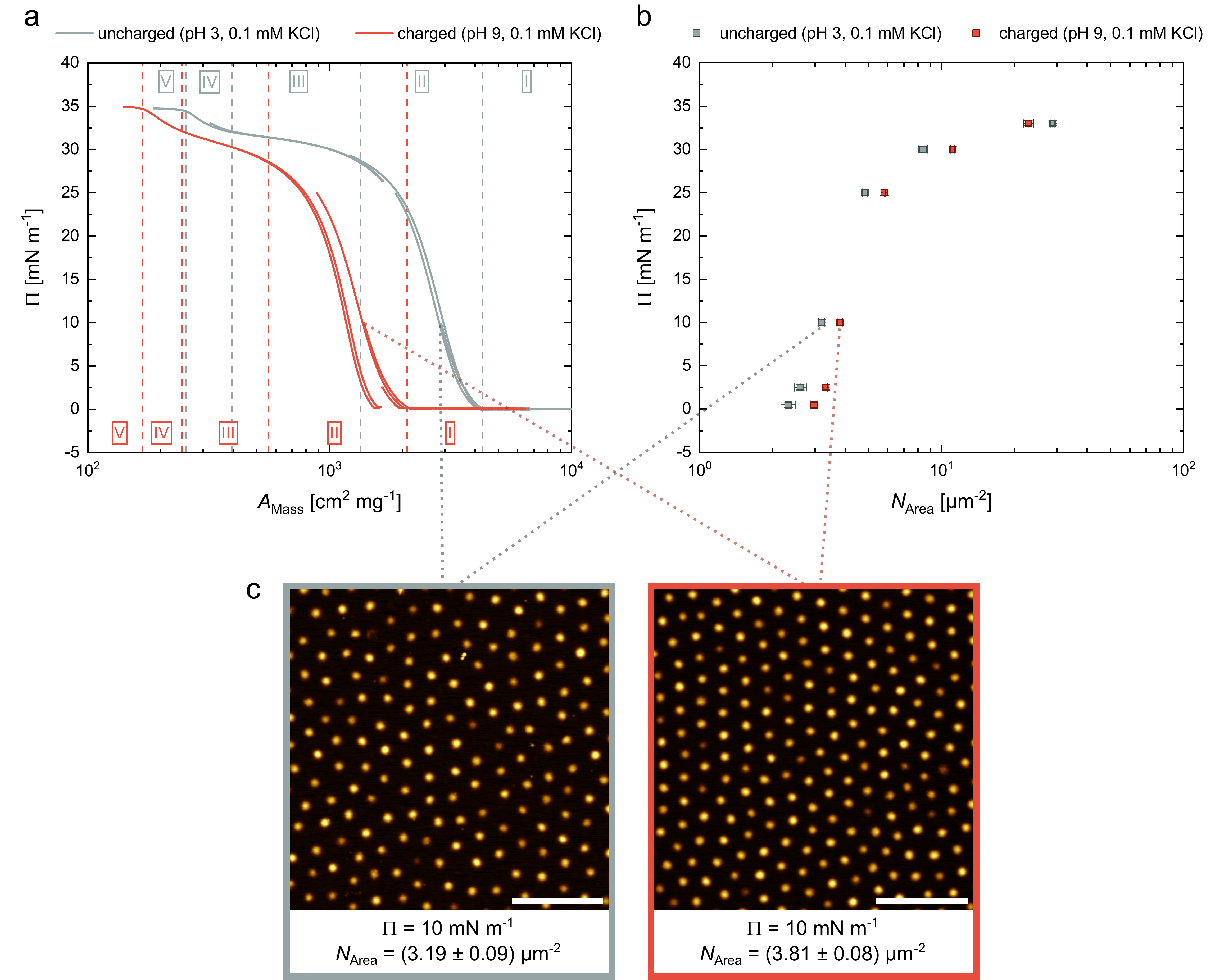}
    \caption{(a) Normalized compression isotherms, surface pressure $\Pi$ \textit{versus} area per mass $A_{\text{Mass}}$, of the S-MAA microgels in the uncharged and charged state. All experiments were conducted at 20~$^\circ$C. Isotherms were normalized to the mass of microgels initially added to the trough. Every line corresponds to an individual measurement. Dashed lines were added to distinguish between different regimes of the compression isotherms labeled by roman numerals. (b) Surface pressure $\Pi$ \textit{versus} number of microgels per area $N_{\text{Area}}$ of the S-MAA microgels in the uncharged and charged state. (c) AFM height images of dried monolayers of the S-MAA microgels in the uncharged (grey box) and charged state (red box) deposited at $\Pi$~=~10~mN~m\textsuperscript{-1}. Scale bars are equal to 2~$\upmu$m.}
    \label{S-MAA_Isotherms_Images_Depositions}
\end{figure}

To clarify the role of electrostatics on the interfacial properties of polyelectrolyte microgels, we start by \textit{reconsidering the compression experiments} performed by Geisel \textit{et al.} \cite{Geisel2014_2} Therefore, we remeasured the compression isotherms of the S-MAA microgels in the uncharged and charged state. They are included in Figure~\ref{S-MAA_Isotherms_Images_Depositions}a.

The data plotted show the same behavior that has been reported by Geisel \textit{et al.} \cite{Geisel2014_2} The onset of the surface pressure increase in isotherms of the charged state is shifted towards lower values of area per mass $A_{\text{Mass}}$ compared to isotherms of the uncharged state, suggesting that charged microgels are easier compressible than uncharged ones. The general shape of the isotherms, i.e., the evolution of surface pressure upon compression of the monolayer, is qualitatively the same for both charge densities and can be divided into five characteristic regimes, labeled (I) to (V). The regimes have been described in the literature. \cite{Pinaud2014, Geisel2014, Geisel2015, Picard2017, Rey2016, Rey2017, Scheidegger2017} They correspond to: (I) a diluted state in which $\Pi$~=~0~mN~m\textsuperscript{-1}, (II) microgels in contact and compression of their coronae resulting in a first sharp increase in surface pressure, (III) a plateau region associated with an isostructural phase transition between two hexagonal lattices, (IV) compression of the microgel cores leading to a second increase in surface pressure, and (V) collapse of the monolayer. The maximum surface pressure reached before the monolayer fails does not depend on the charge density and is approximately 35~mN~m\textsuperscript{-1}.

Although the normalized isotherms of the uncharged and charged state show a significant difference in $A_{\text{Mass}}$ at which the surface pressure first starts to increase, we have to consider how the data are normalized. Normalization of the area to the mass of microgels that are initially added to the Langmuir trough is commonly done for these types of experiments. \cite{Geisel2014, Geisel2015, Pinaud2014, Picard2017} It is assumed that all of the added microgels adsorb to the interface and that they remain there during compression of the monolayer. However, if for some reason, the actual interfacial number concentration is lower, normalization to the mass is flawed and will intrinsically cause a shift of the isotherms to smaller values of $A_{\text{Mass}}$.

We can think of two possibilities that may affect the interfacial number concentration: microgels can desorb from the interface or not all of the initially added microgels adsorb to the interface in the first place. Desorption of microgels is unlikely as their adsorption energy is comparable to values found for solid particles ($\approx$~10\textsuperscript{6}~$k_{\text{B}}T$), which are considered to be irreversibly attached. \cite{Monteillet2014} Incomplete adsorption, on the other hand, may already be visible in Figure \ref{S-MAA_Isotherms_Images_Depositions}a. In some of the individual isotherms measured for the charged state (red lines), the surface pressure starts to rise at $A_{\text{Mass}}$~$\approx$~2000~cm\textsuperscript{2}~mg\textsuperscript{-1}, while for others, it increases at $A_{\text{Mass}}$~$\approx$~1500~cm\textsuperscript{2}~mg\textsuperscript{-1}. For the latter, four times the initial amount was used than for the other measurements. It is reasonable that the number of microgels not adsorbing is higher, the greater the mass initially added, thus, causing a more pronounced shift of the isotherm to lower $A_{\text{Mass}}$. In the uncharged state (grey lines), incomplete adsorption is not a problem, as isotherms of measurements with different initial masses all superimpose and form one consistent master curve. Therefore, charged microgels seem to be less interfacial active or adsorb worse to oil-water interfaces than uncharged ones.

To rule out that the different onset point observed for the isotherms of the charged state compared to the ones of the uncharged state is not the result of a potentially wrong normalization of the area to the mass, we need to determine the actual number of microgels at the interface. Hence, we deposited monolayers at controlled surface pressures and visualized them in dry state. Figure \ref{S-MAA_Isotherms_Images_Depositions}c includes AFM height images obtained at $\Pi$~=~10~mN~m\textsuperscript{-1} as an example, images for other surface pressures can be found in Figure S4. Applying quantitative analysis to each micrograph allows for calculation of the number of microgels per area $N_{\text{Area}}$. The data are listed in Table S1 and the dependency of the surface pressure on $N_{\text{Area}}$ is plotted in Figure \ref{S-MAA_Isotherms_Images_Depositions}b. 

The difference in the values, here of $N_{\text{Area}}$, at which we first register an increase in surface pressure for the uncharged and charged state persists in this representation of the isotherms. However, the magnitude is significantly smaller than in Figure \ref{S-MAA_Isotherms_Images_Depositions}a. At the same surface pressure, we localize more microgels per area in the charged than in the uncharged state. The monolayer of charged microgels needs to be compressed more before the same surface pressure is registered. Put in another way, at the same interfacial concentration, the surface pressure measured for the charged microgels is lower than for the uncharged ones. Only at high compression states, i.e., large surface pressures, the behavior of $N_{\text{Area}}$ is the opposite, and more microgels are localized in the uncharged than in the charged state.

\begin{figure}[H]
    \centering
    \includegraphics[scale=0.35]{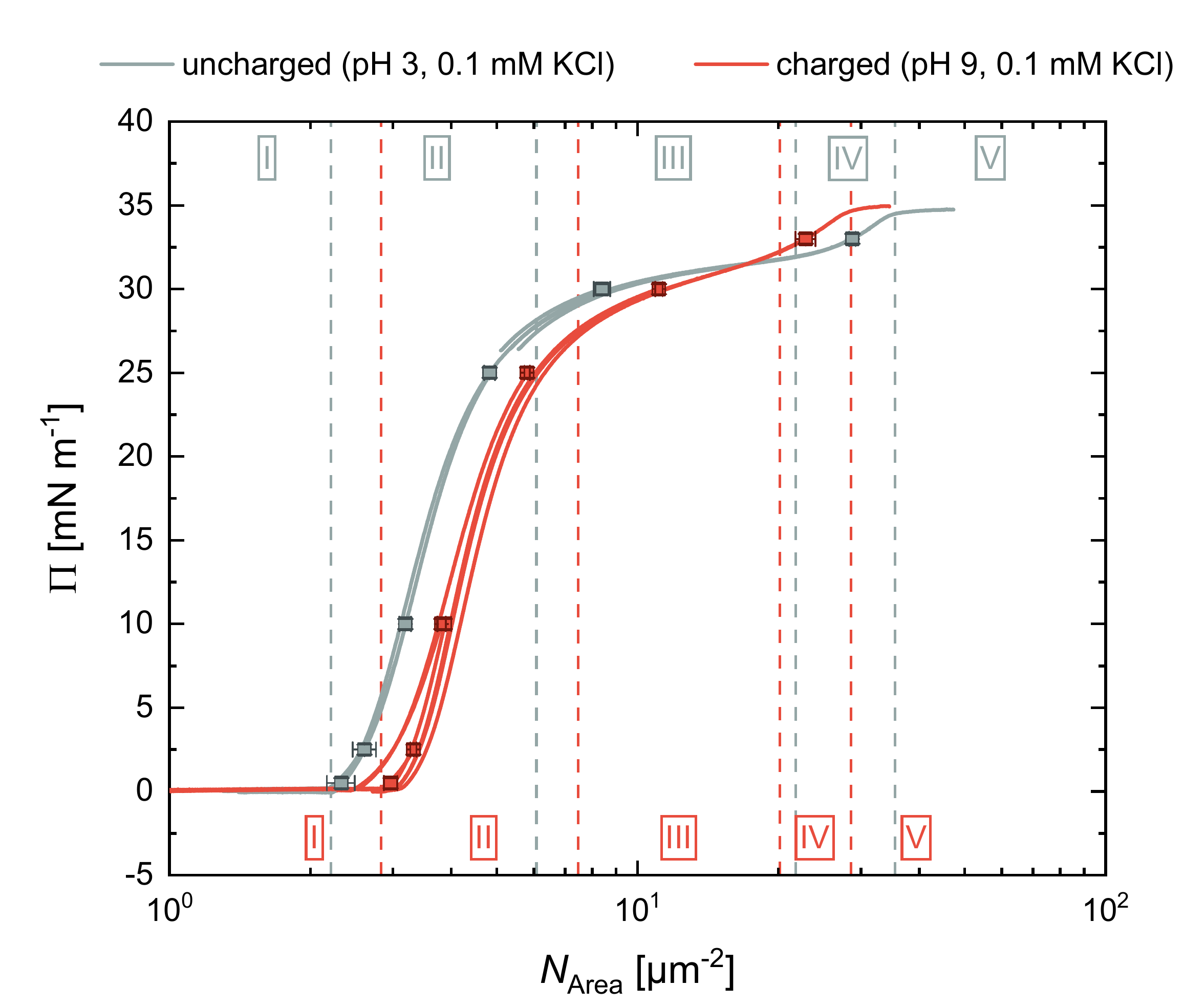}
    \caption{Renormalized compression isotherms, surface pressure $\Pi$ \textit{versus} microgels per area $N_{\text{Area}}$, of the S-MAA microgels in the uncharged and charged state. Every line corresponds to an individual measurement. Dashed lines were added to distinguish between different regimes of the compression isotherms labeled by roman numerals. Data obtained from deposition experiments are included as dots.}
    \label{S-MAA_Isotherms_Renormalized}
\end{figure}

With access to $N_{\text{Area}}$, we can \textit{renormalize the compression isotherms} in Figure \ref{S-MAA_Isotherms_Images_Depositions}a. This allows us to include the data obtained from compression and deposition experiments within the same plot, but without the need to assume that all of the microgels added to the trough adsorb to the interface. The renormalized isotherms are included in Figure \ref{S-MAA_Isotherms_Renormalized}.

With Figure \ref{S-MAA_Isotherms_Renormalized}, we show that microgels in the charged state initially need to be compressed more to reach the same surface pressure as in the uncharged state. Strikingly, in the phase transition region, regime (III), the isotherms of both charge states cross, and the phase transition regime itself is shorter in the isotherms of charged microgels compared to ones of uncharged microgels. After the phase transition is completed, the uncharged microgels need to be compressed more to reach equal surface pressures. Our data clearly demonstrate that electrostatics do affect the interfacial properties of polyelectrolyte microgels.

\subsubsection{Phase behavior at oil-water interfaces}

With the correctly normalized isotherms at hand, we now evaluate the impact of charges on the phase behavior of microgels at oil-water interfaces in more detail. The two-dimensional phase behavior is directly linked to the microgels' interfacial structure. \cite{Pinaud2014, Bochenek2019, Rey2016, Rey2017, Scheidegger2017, Scotti2019} Hence, before discussing the isotherms any further, it is worth taking a look at the morphology of adsorbed microgels in the dilute state where they are not in contact with each other. To that end, we added a relatively small amount of the S-MAA microgels to the oil-water interface, so that the surface pressure remained at 0~mN~m\textsuperscript{-1}, and deposited them without compression of the monolayer. AFM phase images, embedded in Figure S5a, show that uncharged and charged microgels both display the characteristic "fried-egg"-like structure with an inner core and an outer corona region. Determining the diameter probability functions from the images, Figure S5b, reveals that the lateral dimensions, i.e., the in-plane diameter of the full microgels as well as only the core regions, are nearly identical for both charge states. However, it has to be mentioned that the AFM micrographs are of dried microgels that could have undergone transitions in their size and/or morphology upon deposition and subsequent drying. Nevertheless, the absence of a distinct lateral size change with pH is in agreement with results obtained from cryo-scanning electron microscopy. \cite{Geisel2012}

Besides the lateral structure, one has to consider that while being adsorbed to and deformed at the interface, microgels remain three-dimensional objects with a large fraction of the polymer network still situated in the aqueous phase. For temperature-responsive microgels, these fractions retain their stimuli-responsiveness and are deswollen above the volume phase transition temperature (VPTT). \cite{Maestro2018, Harrer2019, Bochenek2019} The temperature-dependent collapse in three dimensions influences their two-dimensional phase behavior. \cite{Bochenek2019} To the best of our knowledge, there are no comparable experimental studies regarding the out-of-plane swelling/deswelling of pH-responsive microgels at the liquid-liquid or liquid-air interface, but microgel films at the solid-liquid interface are thicker in the charged than in the uncharged state. \cite{Howard2010} In addition, computer simulations at the fluid interface demonstrate that the network of the adsorbed microgel in the charged state is significantly more swollen into the polar phase than in the uncharged state. \cite{Gavrilov2019} Therefore, one can expect that in polyelectrolyte microgels, the fractions located in the aqueous side of the interface preserve their pH-responsiveness.

Keeping the two- and three-dimensional morphology of the uncharged and charged microgels in mind, we return to the compression isotherms. As pointed out before, the general shape of the isotherms, i.e., the number and order of regimes in which they can be divided, is the same for both charge states. Yet, at lower compression (before phase transition), charged microgels need to be compressed further than uncharged microgels to reach the same surface pressure, while the behavior reverses at higher compression (after phase transition).

\begin{figure}[H]
    \centering
    \includegraphics[scale=0.35]{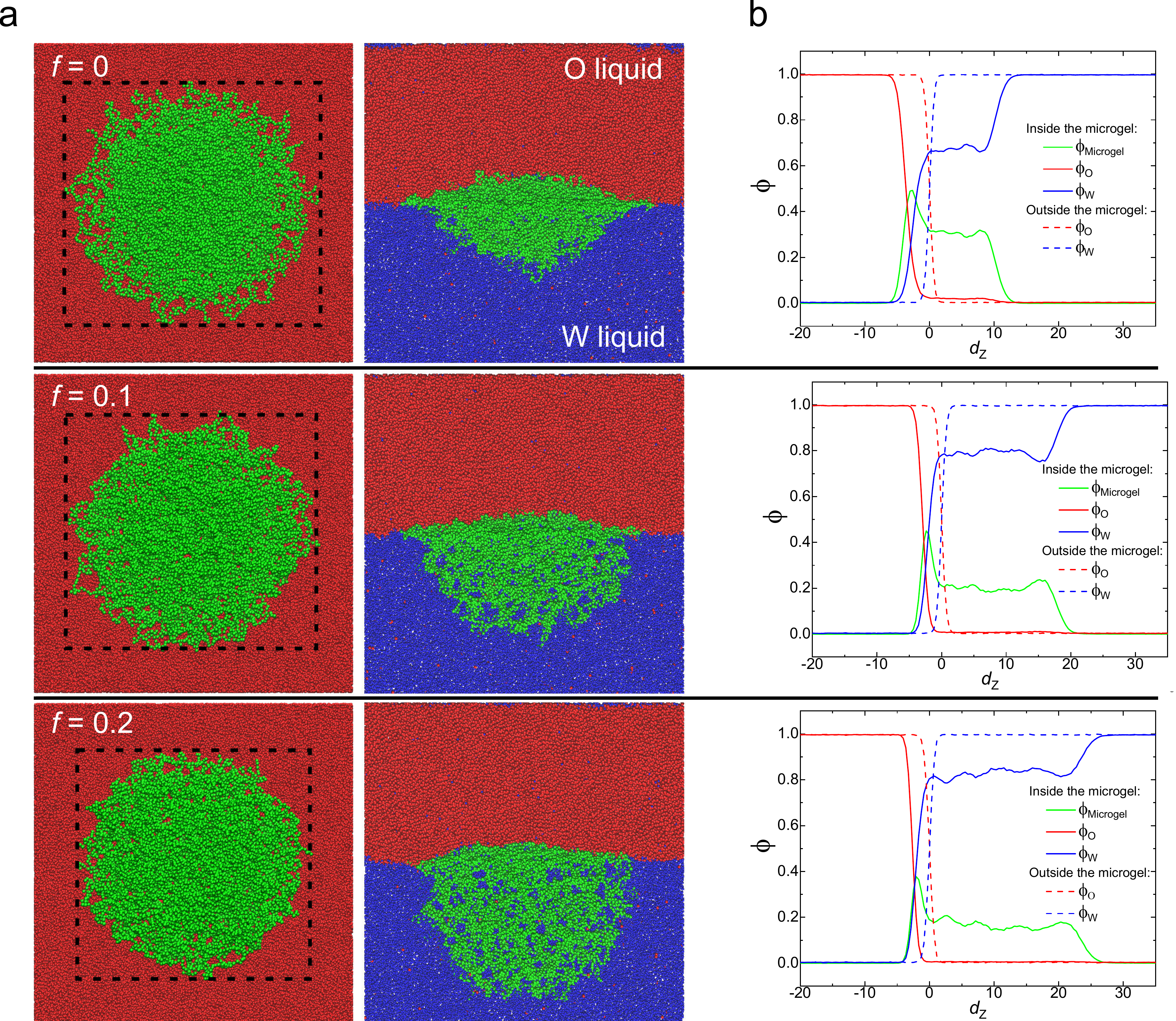}
    \caption{(a) Snapshots (top and side view) of adsorbed microgels for different fractions of charged monomer units $f$ adsorbed at a liquid-liquid interface. The interaction parameter between the uncharged monomer units and the O liquid is $\chi_{\text{uO}}$~=~0.28. (b) Density profiles, volume fraction $\phi$ \textit{versus} distance from the liquid-liquid interface $d_{\text{Z}}$, at the respective conditions.}
    \label{SIM_Snapshots}
\end{figure}

To find an explanation for the different interfacial concentrations between the uncharged and charged state at which the surface pressure first starts to increase, we performed \textit{DPD simulations} of polyelectrolyte microgels at the liquid-liquid interface. In a previous work, the gyration radius was calculated to describe the size of the microgels at different degrees of charging and incompatibility between the uncharged monomer units and the liquids. \cite{Gavrilov2019} While the gyration radius is a convenient tool to characterize the size, it may not reflect certain features of complex shapes, which are crucial for certain scenarios. When we consider the compression of the interface with adsorbed microgels, we are primarily interested in the effective area of the microgels at the interface. This would allow us to assess at what values of interfacial area per microgel $A_{\text{Microgel}}$ they start to "feel" each other upon compression. We can define $A_{\text{Microgel}}$ as the average area of the smallest rectangle that fits the projection of all the microgel monomer units onto the interface. When calculating $A_{\text{Microgel}}$, we considered the case when the less polar solvent O is bad for the uncharged monomer units. The salt-free case was considered for simplicity. Figure \ref{SIM_Snapshots}a presents snapshots (top and side view) of the adsorbed microgels for different fractions of charged monomer units $f$ that were obtained at $\chi_{\text{uO}}$~=~0.28; $\chi_{\text{uO}}$ being the interaction parameter between the uncharged monomer units and the O liquid. The black rectangles show how the effective microgel area was determined. The dependence of $A_{\text{Microgel}}$ on the $\chi_{\text{uO}}$ value for different $f$ is included in Figure S11.

We see that if the value of $\chi_{\text{uO}}$ is not too large, the area of charged microgels is indeed smaller than that of uncharged microgels. Such behavior can be explained by more effective adsorption of the uncharged microgels, i.e., they are flattened more to prevent unfavorable contacts between the two liquids, while the presence of residual charge and counterion osmotic pressure in the charged microgel prevent it from having a too-small thickness. However, this is somewhat surprising given that in previous works it was shown, that the in-plane components of the gyration radius practically did not change upon increasing the value of $f$. The reason for such behavior is in fact rather simple: there are more subchains, i.e., monomer units, located closer to the center of mass of the adsorbed uncharged microgel compared to the charged one, in which the radial density distribution is more homogeneous. As can be seen in Figure \ref{SIM_Snapshots}a, there is a discernible region with low density at the periphery of the uncharged microgel (despite the simulated microgels having ideal topology and, therefore, no corona), while no such region is observed for the charged microgel.

Furthermore, in Figure \ref{SIM_Snapshots}b, we include the density profiles at the respective conditions. The density profiles inside the microgel were averaged using the microgel central part, where its shape is almost independent of the in-plane coordinates. For $f$~=~0, we see that the microgel swells into the W phase. In agreement with previous studies, the microgel is still partly located in the O phase, but due to the non-zero $\chi_{\text{uO}}$, the O liquid inside the microgel is mostly replaced by the W liquid. This essentially causes a slight shift of the liquid-liquid interface inside the microgel into the O phase. \cite{Geisel2012, Geisel2015_2, Rumyantsev2016} Introducing charges in the microgel at $f$~=~0.1 and 0.2 leads to further swelling into the W phase. Again, the microgel, even its part located in the O phase, is mostly filled with W liquid. It is worth noting that while the microgel is significantly swollen into the W liquid, it still facilitates the mixing of the liquids as the volume fraction of the O liquid in the microgel is higher than in the bulk W liquid. The effect, as expected, is less pronounced for higher values of $f$ due to the lower polymer volume fraction inside more swollen microgels as well as the incompatibility between the charged monomer units and the O liquid.

Based on the data obtained from DPD simulations, we conclude that the different onset points of the surface pressure increase observed in isotherms of the uncharged and charged state, are related to a charge-dependent difference in in-plane interactions of the respective microgels with the phases comprising the interface. Monolayers of charged microgels need to be compressed further, simply because charged microgels are characterized by a smaller in-plane effective area over uncharged microgels, despite being more swollen in three dimensions.

Once the microgels are in contact with each other, they form a hexagonal lattice, and the work exerted on the monolayer is associated with compression of the coronae. From Figure \ref{S-MAA_Isotherms_Renormalized}, we see that in regime (II), the slope with which the surface pressure increases is the same (within the variations of the experiment). This implies that the force required to compress the microgels in this regime is not affected by electrostatic interactions. Still, the shift of the isotherms of the charged microgels to higher values of $N_{\text{Area}}$ remains due to their smaller effective area.

When the center-to-center distance between microgels in the monolayer is equal to the size of the core, the coronae are fully compressed, and the \textit{isostructural phase transition} starts. \cite{Bochenek2019} In this regime, the first hexagonal lattice melts, while a second hexagonal lattice with smaller lattice constant forms. \cite{Rey2016} Here, we see that in regime (III), the isotherms of the uncharged and charged state cross. Furthermore, regime (III) is shorter for the charged state, which suggests that the transition of the microgels into the second hexagonal lattice is completed earlier in the monolayers of the charged microgels than in the ones of uncharged microgels. 

From earlier work, we learned that for temperature-responsive microgels, the out-of-plane fractions of the network influence the compressibility of the cores. Increasing the temperature above the VPTT causes a collapse of the parts situated in the aqueous phase and leads the core to be denser and thinner and, therefore, less compressible than in the non-collapsed state. \cite{Bochenek2019} Thus, when the core is dominating the microgel-to-microgel interactions within the monolayer, the additional out-of-plane contribution needs to be taken into consideration.

As pointed out before and seen in the snapshots in Figure \ref{SIM_Snapshots}, the out-of-plane portions of polyelectrolyte microgels retain their pH-responsiveness. These fractions are more swollen in the charged state than in the uncharged one. Hence, charged microgels start to interact with each other sooner in three dimensions due to electrostatic repulsion. The lattice constant of the second hexagonal phase should be larger in monolayers of the charged state compared to the uncharged state. This can explain a shorter phase transition regime and the start of regime (IV) at lower values of $N_{\text{Area}}$. With our data, we can compare the lattice constants in monolayers of uncharged and charged microgels before and after phase transition. Before, the center-to-center distance at a surface pressure of $\Pi$~=~25~mN~m\textsuperscript{-1} is smaller in the charged state ($d_{\text{cc,1}}$~$\approx$~430~$\pm$~20)~nm) than in the uncharged state ($d_{\text{cc,1}}$~$\approx$~(450~$\pm$~35)~nm). At a surface pressure of $\Pi$~=~33~mN~m\textsuperscript{-1}, where phase transition is completed and all microgels are in the second hexagonal phase, the behavior reverses and $d_{\text{cc}}$ is larger in the charged state ($d_{\text{cc,2}}$~$\approx$~(190~$\pm$~20)~nm) compared to the uncharged one ($d_{\text{cc}}$~$\approx$~(160~$\pm$~20)~nm). Therefore, at high compression, i.e., after phase transition, the compressibility of the monolayer is determined by out-of-plane interactions between the microgel portions situated farther into the aqueous side of the interface. Consequently, monolayers of charged microgels can be compressed less because the difference in lattice constants between the first and second hexagonal phase is smaller for the charged than for the uncharged state. 

Strikingly, the maximum surface pressure registered before failure, regime (V), is nearly identical for both charge states.

\subsubsection{Ordering within the monolayer}

Besides the number of microgels per area and the center-to-center, quantitative image analysis of our AFM images provides information on the ordering of microgels within the deposited monolayers as well. Recent computer simulations of polyelectrolyte microgels at the liquid-liquid interface studied the role of charges on the ordering in monolayers of microgels. \cite{Gavrilov2019} The introduction of charges within the networks and, thus, long-range repulsive forces, led to an almost ideal hexagonal packing of the microgels even before they were in contact with each other, i.e., at distances between them larger than their interfacial diameter. The positions were random when the simulated microgels did not contain any charged fractions. In our experiments, we can characterize the ordering with two parameters. The short-range order can be described by the hexagonal order parameter $\Psi_6$. A perfectly ordered hexagonal lattice has an order parameter of 1, while in a disordered structure, it tends to 0. Information on the long-range ordering is provided by the radial distribution function $g(r)$.

The AFM images of the depositions performed in the dilute state at $\Pi$~=~0~mN~m\textsuperscript{-1}, Figure S5a, show no clear indication of a better ordering. However, simulations were performed under salt-free conditions, wherein the experiments we always kept a low background concentration of salt to ensure the reproducibility of the measurements. The presence of salt ions can lead to screening effects that hinder the formation of long-range ordering.

Nevertheless, we do find an influence of the charge state on the ordering when the surface pressure starts to increase and microgels get into contact with each other. The values for $\Psi_6$ are listed in Table S1 and plotted in Figure S6. The general progression for $\Psi_6$ with increasing $N_{\text{Area}}$ is qualitatively the same for both charge states and is linked to the different regimes of the compression isotherms. \cite{Bochenek2019, Rey2016, Rey2017, Scheidegger2017} The values of $\Psi_6$ are significantly higher in monolayers of the charged microgels than in the ones of the uncharged microgels for all investigated surface pressures.

\begin{figure}[H]
    \centering
    \includegraphics[scale=0.35]{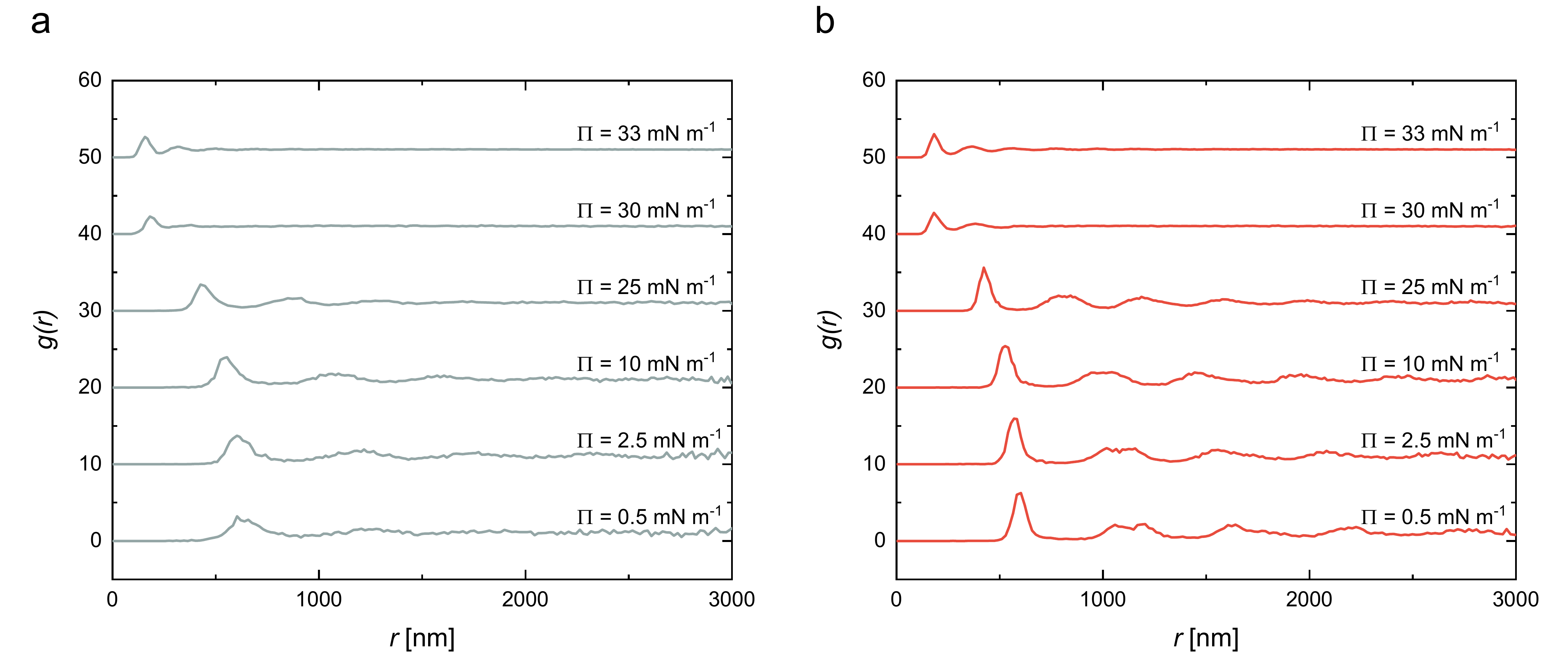}
    \caption{Radial distribution functions, $g(r)$ \textit{versus} distance $r$, in dried monolayers of the S-MAA microgels in the (a) uncharged and (b) charged state deposited at controlled surface pressures $\Pi$.}
    \label{S-MAA_Radial_Distribution_Functions}
\end{figure}

Furthermore, the better ordering in monolayers of the charged microgels is also deducible from the radial distribution functions depicted in Figure \ref{S-MAA_Radial_Distribution_Functions}.

Up to a surface pressure of $\Pi$~=~25~mN~m\textsuperscript{-1}, the functions plotted for the charged state show much more defined peaks for the first maximum, which corresponds to the center-to-center distance, as well as the other positions characteristic for a hexagonally ordered lattice ($2 d_{\text{cc}}$ and $\sqrt{3 d_{\text{cc}}^2}$). At $\Pi$~=~30~mN~m\textsuperscript{-1}, the first peak decreases and a second peak at a lower distance emerges. In this regime, the phase transition takes place and the long-range translational order fails. The peaks correspond again to $d_{\text{cc}}$ and $2 d_{\textsubscript{cc}}$ once phase transition is completed, as can be seen for $\Pi$~=~33~mN~m\textsuperscript{-1}. 

\subsubsection{Influence of size}

\begin{figure}[H]
    \centering
    \includegraphics[scale=0.35]{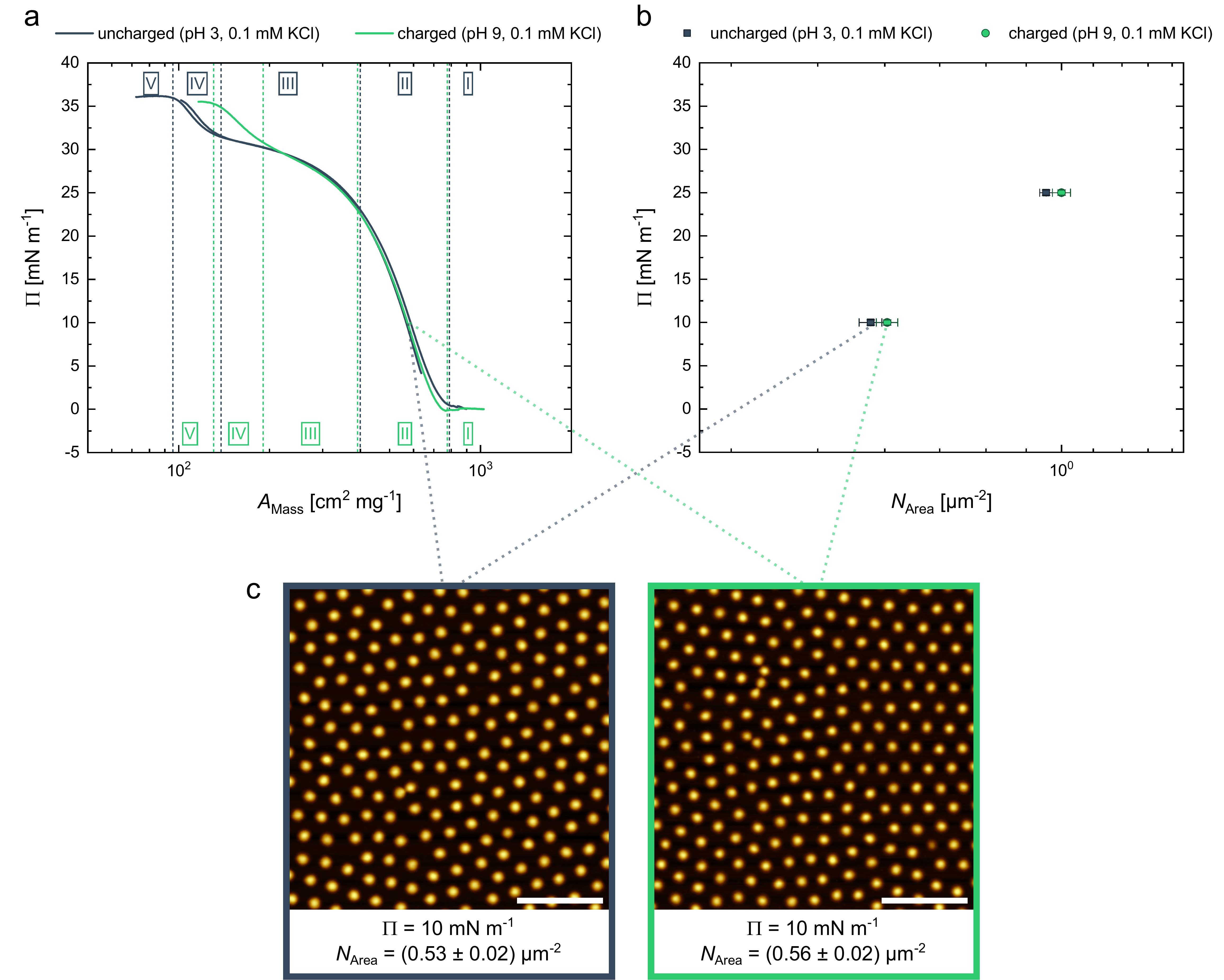}
    \caption{(a) Normalized compression isotherms, surface pressure $\Pi$ \textit{versus} area per mass $A_{\text{Mass}}$, of the L-MAA microgels in the uncharged and charged state. All experiments were conducted at 20~$^\circ$C. Isotherms were normalized to the mass of microgels initially added to the trough. Every line corresponds to an individual measurement. Dashed lines were added to distinguish between different regimes of the compression isotherms labeled by roman numerals. (b) Surface pressure $\Pi$ \textit{versus} number of microgels per area $N_{\text{Area}}$ of the L-MAA microgels in the uncharged and charged state. (c) AFM height images of dried monolayers of the L-MAA microgels in the uncharged (dark grey box) and charged state (green box) deposited at $\Pi$~=~10~mN~m\textsuperscript{-1}. Scale bars are equal to 5~$\upmu$m.}
    \label{L-MAA_Isotherms_Images_Depositions}
\end{figure}

Having established that there is an electrostatic contribution to the compressibility of the S-MAA microgels, we search for an explanation why in the work of Picard \textit{et al.}, the isotherms showed no variation with the ionization state of the microgels. \cite{Picard2017} A substantial difference between their work and our work (or the work of Geisel \textit{et al.}) is the size of the investigated system. The microgels employed by Picard \textit{et al.} are significantly larger with a size of $R_{\text{h}}$~$>$~500~nm when charged. The size can affect the two-dimensional phase behavior of microgels, e.g., large microgels can form clusters at the interface even in the dilute state due to attractive capillary forces. \cite{Scheidegger2017}. Therefore, we performed the same type of compression and deposition experiments carried out for the S-MAA microgels with the larger L-MAA microgels. The latter ones have dimensions similar to the microgels utilized by Picard \textit{et al.} Figure \ref{L-MAA_Isotherms_Images_Depositions}a includes the obtained compression isotherms.

The general shape of the compression isotherms for both charge densities is the same and can be divided into the five characteristic regimes. Strikingly, the isotherms of the larger L-MAA microgels do not show a difference in the onset point of the surface pressure increase between the uncharged and the charged state, which is in perfect agreement with the data reported by Picard \textit{et al.} \cite{Picard2017} However, the isotherms superimpose only until the isostructural phase transition takes place. In analogy to what is observed for the S-MAA microgels, the phase transition regime is shorter for the charged state, and the crossing from regime (III) to regime (IV) occurs at higher values of $A_{\text{Mass}}$. The same applies to the transition from regime (IV) to (V), although the maximum surface pressure reached before the collapse of the monolayer is nearly identical for both charge states. Apparently, charges have no influence on the contact point of large microgels or their compressibility in the early stages of the compression isotherm, regime (II). However, at higher compression, charges need to be considered, where they seem to have the same effect on the compressibility of the large L-MAA microgels as determined for the smaller S-MAA microgels.

We deposited monolayers at surface pressures of $\Pi$~=~10~mN~m\textsuperscript{-1} and $\Pi$~=~25~mN~m\textsuperscript{-1} and analyzed the microstructure. This way, we can examine if the interfacial number concentration is the same for both charge densities in the regions where the isotherms are superimposing. AFM height images of dried monolayers at $\Pi$~=~10~mN~m\textsuperscript{-1} are included in Figure \ref{L-MAA_Isotherms_Images_Depositions}c, the others in Figure S7. The results from the quantitative analysis are listed in Table S2. 

Plotting $N_{\text{Area}}$ in dependency of $\Pi$ shows that the interfacial number concentration is almost identical at the same surface pressure, Figure \ref{L-MAA_Isotherms_Images_Depositions}c. In the images obtained at $\Pi$~=~25~mN~m\textsuperscript{-1}, we see that phase transition is already taking place. Although the isotherms are still superimposing at that point and $N_{\text{Area}}$ is almost the same, the center-to-center distance of microgels in the second phase is larger for the charged state ($d_{\text{cc,2}}$~$\approx$~(820~$\pm$~75)~nm) than for the uncharged state ($d_{\text{cc,2}}$~$\approx$~(770~$\pm$~65)~nm). The data confirm what we stated earlier for the S-MAA microgels. The lattice constant of the second phase is larger for the charged microgels because of electrostatic interactions between the more swollen microgel fractions in the aqueous phase. Therefore, the phase transition is completed earlier.

Radial distribution functions are in included in Figure S8. Monolayers of both charge states are characterized by a high degree of ordering as the peaks are well pronounced independent of the microgels' charge state.

The fact that in isotherms of the large L-MAA microgels, the concentration at the onset of the surface pressure increase does not depend on the charge state can be explained by the following.  As we know from the computer simulations, the difference in the onset points is related to a smaller in-plane effective area of charged microgels over uncharged ones. However, the growth of the microgel size diminishes its deformation at the interface. Indeed, adsorption and increase of the contact area is caused by a gain in the interfacial energy, which is proportional to $L^2$; $L$ being the linear dimension of the microgel. This flattening is opposed by the elasticity of the network and electrostatics (in case of the charged microgels). All these contributions to the free energy of the microgel are proportional to the microgel volume, i.e., to $L^3$. Therefore, an increase in the microgel size leads to the dominance of volume interactions over the interfacial ones ($L^3$ \textit{versus} $L^2$). It has to be mentioned that macroscopic objects (the thermodynamic limit at $L\to\infty$) cannot be deformed at the interface.

\subsubsection{Influence of the amount of charged groups and salt concentration} 

\begin{figure}[H]
     \centering
     \includegraphics[scale=0.35]{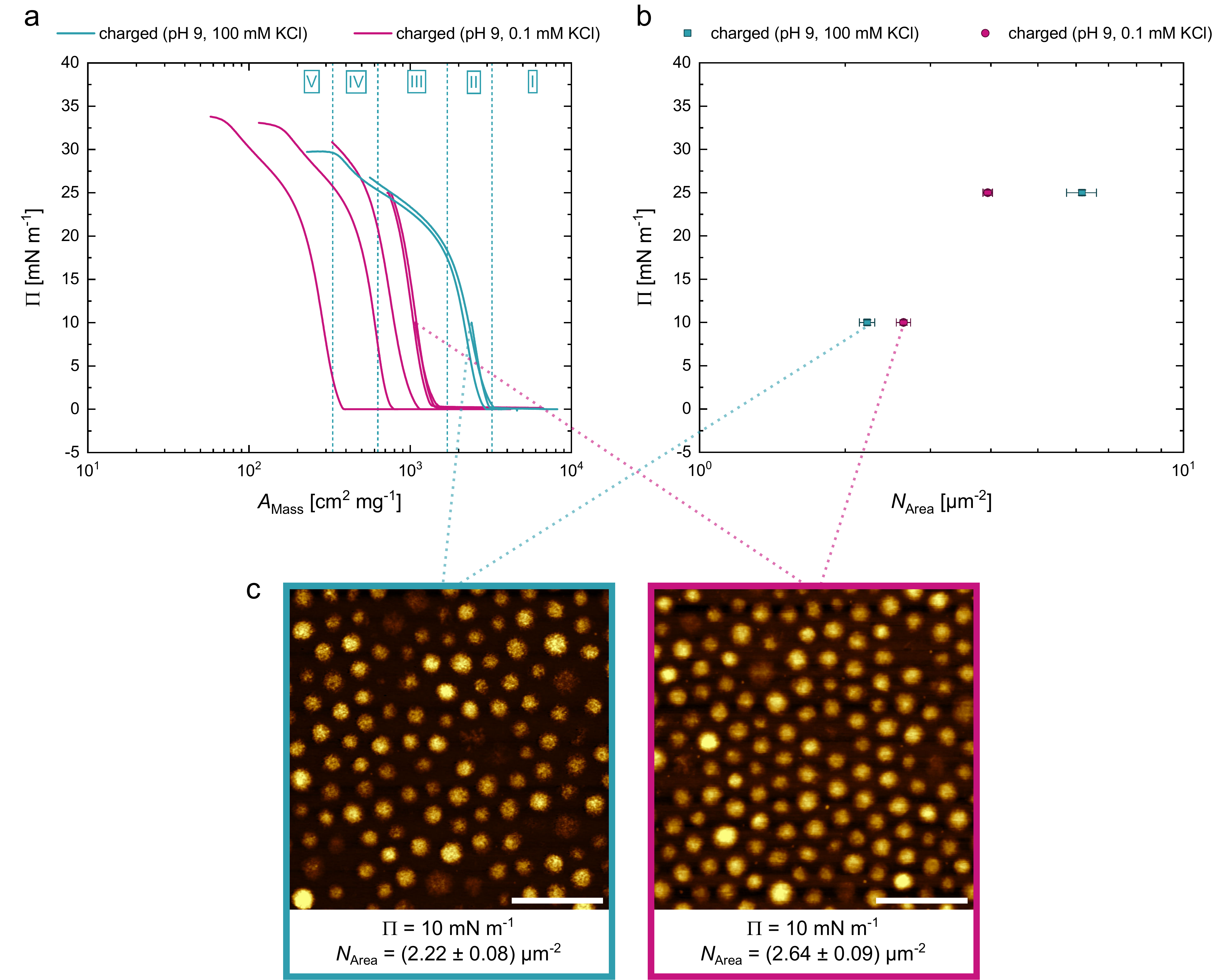}
     \caption{(a) Normalized compression isotherms, surface pressure $\Pi$ \textit{versus} area per mass $A_{\text{Mass}}$, of the HC-IA microgels in the charged state at 100~mM~KCl and 0.1~mM~KCl. All experiments were conducted at 20~$^\circ$C. Isotherms were normalized to the mass of microgels initially added to the trough. Every line corresponds to an individual measurement. Dashed lines were added to distinguish between different regimes of the compression isotherms labeled by roman numerals. (b) Surface pressure $\Pi$ \textit{versus} number of microgels per area $N_{\text{Area}}$ of the HC-IA microgels in the charged state at 100~mM~KCl and 0.1~mM~KCl. (c) AFM height images of dried monolayers of the HC-IA microgels in the charged state at 100~mM~KCl (blue box) and 0.1~mM~KCl (violet box) deposited at $\Pi$~=~10~mN~m\textsuperscript{-1}. Scale bars are equal to 2~$\upmu$m.}
     \label{HC-IA_Isotherms_Images_Depositions}
 \end{figure}

Before we conclude our study, we probe if increasing the amount of charged groups within the polymer network of the microgels can amplify or alter charge effects at the interface. Furthermore, we explore the influence of ionic strength on the interfacial properties. Adjusting the ionic strength by the addition of salt provides another possibility to modulate electrostatic interactions. Therefore, we performed compression and deposition experiments with the \textit{highly charged HC-IA microgels} in their charged state at pH~9, but at different salt concentrations present in the aqueous subphase. The bulk dimensions of the HC-IA microgels under these conditions are in a similar range as determined for the S-MAA microgels. The obtained compression isotherms are shown in Figure~\ref{HC-IA_Isotherms_Images_Depositions}a. 

In the isotherms recorded under conditions where electrostatic interactions are stronger, here at the lower salt concentration of 0.1~mM KCl, the onset of the increase in surface pressure is shifted to smaller values of $A_{\text{Mass}}$ compared to the isotherms acquired at high ionic strength with significant charge screening. This is in agreement with what has been observed for the S-MAA microgels. 

Furthermore, measurements performed at 0.1~mM KCl but with different initially added mass of microgel (violet lines) do not superimpose. The offset between individual measurements is more pronounced than the offset seen for the isotherms of the charged S-MAA microgels in Figure \ref{S-MAA_Isotherms_Images_Depositions}a. As argued before, the lack of a master curve can be related to not all of the initially added charged microgels adsorbing to the interface. It is not surprising that this problem becomes more dominant, the more charges the microgels carry and, thus, the higher their hydrophilicity is. The increased number of charges also influences the phase transition regime. In the violet curves, a distinct plateau region is no longer distinguishable as compared to the isotherms of the charged S-MAA or L-MAA microgels. Therefore, we did not include optical guidelines to differentiate between regimes. The isotherms recorded at high ionic strength where electrostatic interactions are weakened (blue lines) do form a master curve when normalized to the mass. They display a more pronounced phase transition regime and all five characteristic regimes can be assigned.

However, monolayers of the charged HC-IA microgels at different salt concentrations do not fail at the same surface pressure. For 0.1~mM KCl, failure occurs at $\Pi$~$\approx$~35~mN~m\textsuperscript{-1}, while for 100~mM KCl only $\Pi$~$\approx$~30~mN~m\textsuperscript{-1} is reached. The surface pressure is defined as the difference in interfacial tension between the bare interface and the microgel-covered interface. \cite{Moehwald1993} It is known that the interfacial tension of water can change through the addition of salt ions, \cite{Pegram2007} though, the KCl concentrations employed in our experiments are too low to have a significant effect. \cite{Aveyard1976} Therefore, the lower value of $\Pi$ observed at 100~mM KCl compared to 0.1~mM KCl implies a less effective reduction of the interfacial tension by the microgels at higher salt concentrations.

For both salt concentrations, monolayers were deposited at controlled surface pressures and quantitatively analyzed. AFM images are included in Figures \ref{HC-IA_Isotherms_Images_Depositions}c and S9, results from the image analysis in Table S3. Plotting the surface pressure depending on the number of microgels per area, Figure \ref{HC-IA_Isotherms_Images_Depositions}b, shows that modulating electrostatic interactions via ionic strength results in the same trends as compared to modulation via pH. For lower surface pressures, the microgels at lower ionic strength, hence, stronger electrostatic interactions, need to be compressed more than the ones at higher ionic strength. The opposite behavior is observed for larger surface pressures. 

Although the dried HC-IA microgels show some degree of size polydispersity in the AFM height images in Figure \ref{HC-IA_Isotherms_Images_Depositions}c, the ordering in monolayers with stronger electrostatic interactions is still higher compared to the ones with screened charges. Additionally, more pronounced peaks are observed in the radial distribution functions included in Figure S10.

\section{Summary and Conclusions}

In this study, we have investigated the effects of electrostatics on the interfacial properties of polyelectrolyte microgels to clarify the contradictory findings reported so far. Combining compression and deposition experiments with quantitative image analysis and computer simulations, we show that the behavior of pH-responsive microgels adsorbed to oil-water interfaces does depend on the ionization state of the microgels. Compression isotherms of uncharged and charged microgels qualitatively display the same characteristics, and AFM images of deposited monolayers show the same microstructures. However, quantitatively, there are charge- and size-dependent differences in the evolution of surface pressure upon compression.

For smaller microgels, the onset point of the surface pressure increase differs between the uncharged and charged state. Monolayers of charged microgels need to be compressed more than the ones of uncharged microgels before a change in surface pressure is registered. With the help of computer simulations, we relate the difference in the onset point to a smaller in-plane effective area of adsorbed charged microgels as compared to uncharged microgels. Therefore, the compressibility at low compression of the monolayer, i.e., before phase transition, is controlled by in-plane interactions of the respective uncharged or charged microgels with the interface.

Isotherms of the uncharged and charged state intersect in the phase transition regime and at high compression, i.e., after phase transition, charged microgels become less compressible than uncharged ones. Accordingly, the isotherms of the charged state are characterized by a shorter phase transition regime and an earlier failure of the monolayer than the ones of the uncharged state, although the maximum surface pressure reached before collapse is virtually the same. We explain the different compressibility after phase transition with the three-dimensional structure of the microgels. Charged microgels are more swollen perpendicular to the interface than uncharged microgels and, thus, start to interact sooner in three dimensions. This is reflected by a smaller difference between the lattice constants of the first and second hexagonal phase determined for the charged state over the uncharged state. Consequently, at high compression, the compressibility of the monolayer is governed by out-of-plane interactions between the microgel portions further in the aqueous side of the interface.

However, the size of the investigated microgels plays an important role. The onset point of the increase in surface pressure in isotherms of larger microgels does not depend on the presence of charges. For large microgels, the difference in in-plane effective area between uncharged and charged state becomes negligible. The reason for that is a dominance of the volume over the interfacial contribution to the free energy. Yet, at high compression, the isotherms of the uncharged and charged state deviate from each other and show the same behavior as described for the smaller microgels.

Identifying the size of the investigated microgels as a key parameter, we can also explain why the results obtained in the two studies mentioned in the introduction differ. Picard \textit{et al.} analyzed much larger microgels than Geisel \textit{et al.} \cite{Picard2017, Geisel2014_2} Therefore, the size, and not, as suggested by Picard \textit{et al.}, the presence of impurities with pH-dependent interfacial activities, led to the contrasting results between the two works.

In addition to that, we would like to point out that for the right interpretation of the compression isotherms of the S-MAA microgels, it is essential to normalize them to the number of microgels per area instead to the area per mass. The latter assumes that all of the initially added microgels adsorb to the interface, which is found not to be true for the smaller microgels in the charged state and becomes even worse for the highly charged microgels.

Finally, we have shown that monolayers of charged microgels can display a higher degree of ordering than the monolayers of uncharged microgels, which was predicted by computer simulations. \cite{Gavrilov2019} Values found for the hexagonal order parameter for the charged state are larger than for the uncharged state, and the radial distribution functions at comparable surface pressures show better-pronounced peaks.

Our results highlight the much more complex behavior of soft microgels as compared to rigid particles of similar dimensions. Regarding the application of polyelectrolyte microgels as stabilizers for emulsions that can be broken on-demand, we do not recommend the use of large microgels, as their interfacial properties are not affected by electrostatic interactions. Moreover, independent of electrostatics, a systematic study already found large microgels to be less effective emulsion stabilizers than small ones. \cite{Destribats2014} However, even when using small polyelectrolyte microgels, the number of charged groups within the network needs to be considered, as our results clearly show that adsorption of charged microgels to the interface can be limited. In the literature, successful stabilization of emulsions was achieved with microgels containing about 5, but not more than 10~mol\% or wt\%, respectively, of ionizable comonomer. \cite{Brugger2008, Ngai2005, Wiese2013}

\begin{acknowledgement}
The authors thank the Deutsche Forschungsgemeinschaft for financial support within the Collaborative Research Center SFB 985 "Functional Microgels and Microgel Systems" (Project B8). The financial support of the Russian Foundation for Basic Research, Project Nos. 19-03-00472 and 20-33-70242, and the Government of the Russian Federation within Act 211, Contract No. 02.A03.21.0011, is gratefully acknowledged. Part of this research was carried out using the equipment of the shared research facilities of HPC computing resources at Lomonosov Moscow State University.
\end{acknowledgement}

\begin{suppinfo}
HC-IA microgel synthesis (Incorporation of DMI, hydrolysis of DMI, assessment of the purification process); S-MAA microgels at oil-water interfaces (AFM height images, image analysis results, AFM phase images and diameter probability functions, short-range ordering); L-MAA microgels at oil-water interfaces (AFM height images, image analysis results, radial distribution functions); HC-IA microgels at oil-water interfaces (AFM height images, image analysis results, radial distribution functions); Computer simulations of microgels at the liquid-liquid interface (Dependence of the effective microgel area on the interactions parameter)
\end{suppinfo}

\clearpage

\bibliography{references_main}

\clearpage

\begin{figure}[H]
    \centering
    \includegraphics{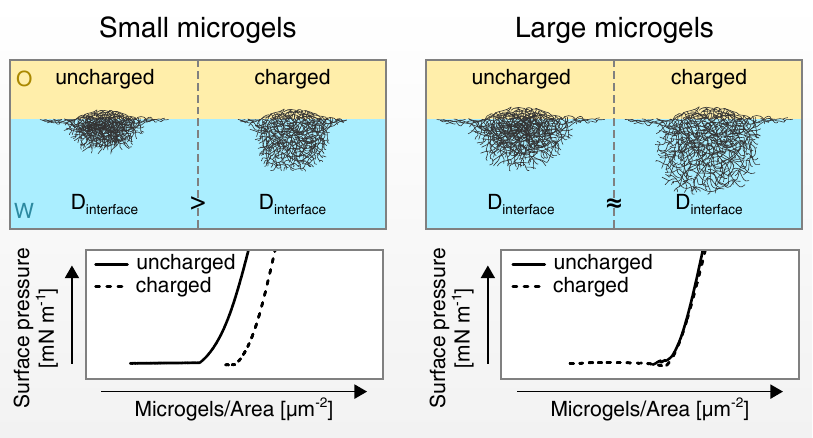}
    \caption{For table of contents only.}
\end{figure}

\end{document}


\clearpage

\tableofcontents

\clearpage

\section*{HC-IA microgel synthesis}

\subsection*{Incorporation of DMI}

\begin{figure}[H]
    \centering
    \includegraphics[scale=0.35]{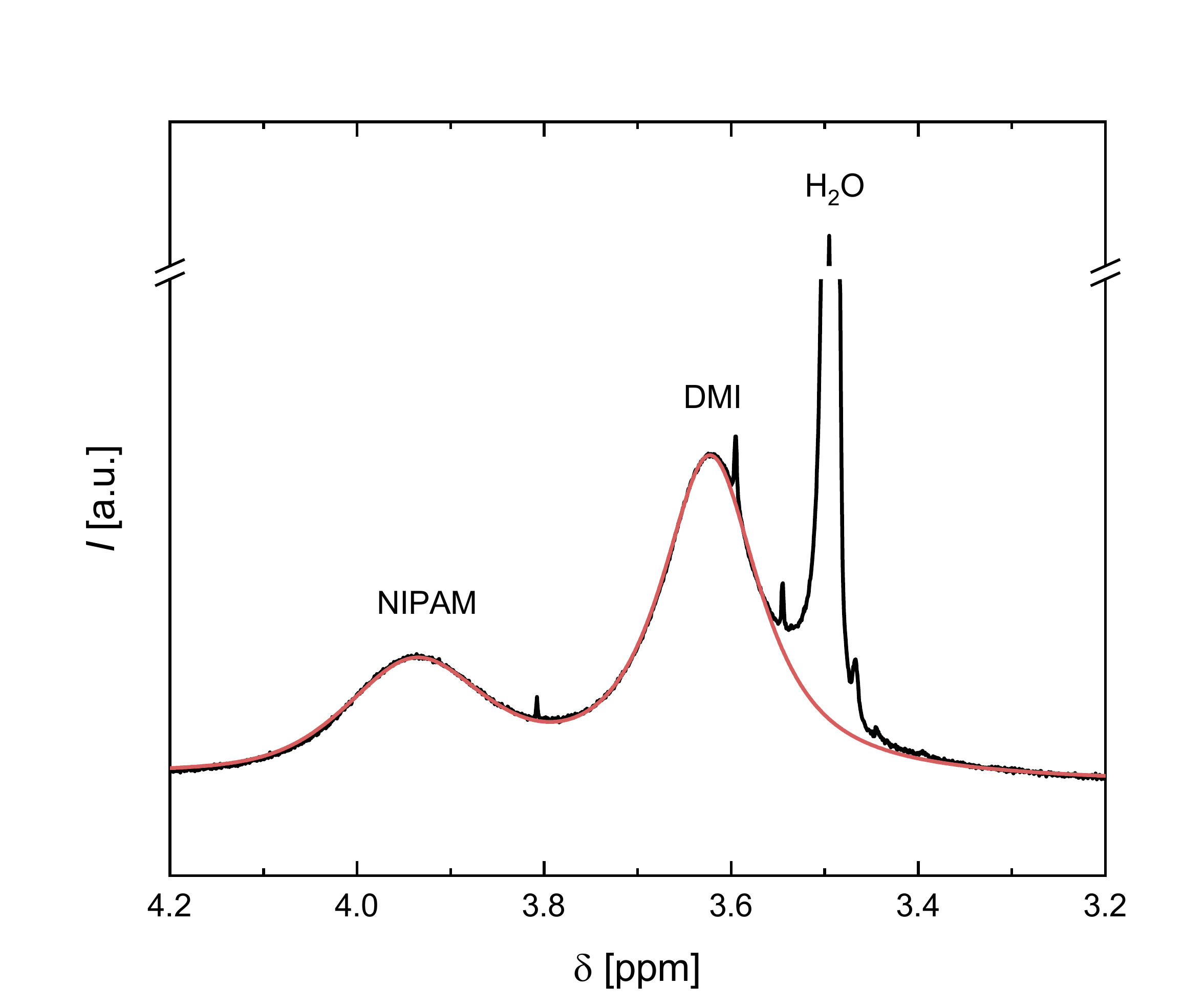}
    \caption{\textsuperscript{1}H-NMR spectrum, Intensity $I$ \textit{versus} chemical shift $\delta$, of the non-hydrolyzed microgels in DMF-d\textsubscript{7}. Phase and baseline were manually corrected.}
    \label{HC-IA_Incorporation}
\end{figure}

The incorporation of DMI into the uncharged precursor microgels was quantified by \textsuperscript{1}H-NMR spectroscopy (Avance III 400 MHz, Bruker Corporation USA). Therefore, a small amount of the non-hydrolyzed microgel solution was lyophilized and subsequently redispersed in deuterated \textit{N, N'}-dimethylformamide (DMF-d\textsubscript{7}). The concentration was 4~mg~mL\textsuperscript{-1}. The obtained spectrum (section of interest) is included in Figure \ref{HC-IA_Incorporation}.

The molar composition of the microgels is determined from a comparison of the integrals of the NIPAM peak ($\delta$~$\approx$~3.94~ppm, 1~H) and the DMI peak ($\delta$~$\approx$~3.62~ppm, 8~H). As a clear boundary between the two peaks is not distinguishable and because the DMI peak is overlapping with the one of residual water, the peaks of NIPAM and DMI were fitted (red line) utilizing the software \textit{dmfit2015}.\textsuperscript{45} With the relative integrals derived from the fit (NIPAM:~0.281, DMI:~0.719), the ratio between protons can be calculated. Thus, the DMI content is found to be 24.2~mol\% (feed: 25.0~mol\%). DMI is almost quantitatively incorporated into the microgels during the first step of the synthesis.

\subsection*{Hydrolysis of DMI}

\begin{figure}[H]
    \centering
    \includegraphics[scale=0.35]{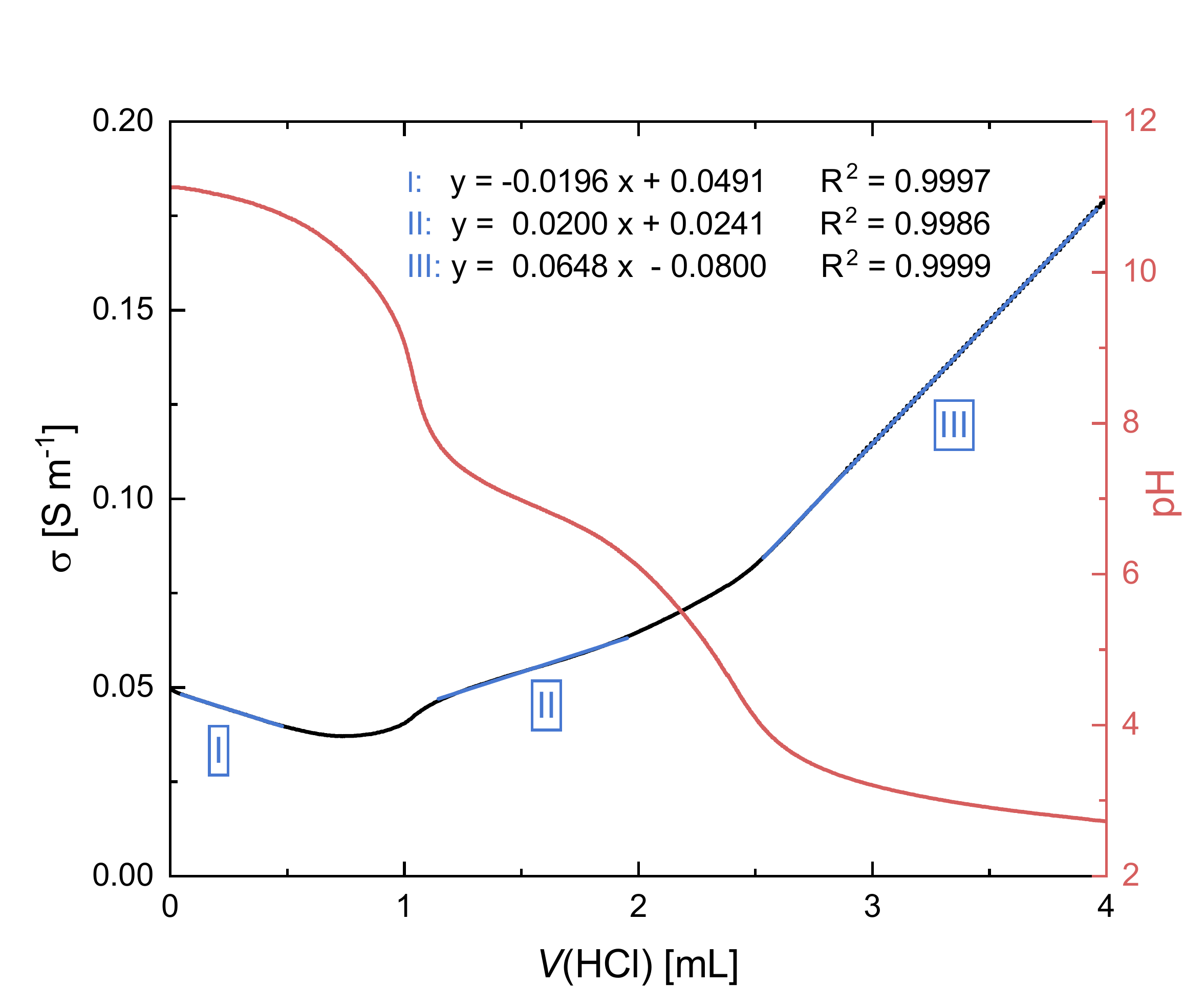}
    \caption{Titration curve, conductivity $\sigma$ and pH \textit{versus} volume of consumed acid $V$(HCl,) of the hydrolyzed microgels.}
    \label{HC-IA_Hydrolysis}
\end{figure}

The extent of hydrolysis, i.e., the conversion of ester groups within the microgels into ionizable carboxyl groups, was quantified utilizing conductometric titration. Therefore, an aqueous solution containing 51.31~mg of the purified hydrolyzed microgels was titrated with 0.1~M HCl starting at pH~11. Conductivity $\sigma$ and pH were recorded depending on the volume of added acid at 20~$^\circ$C. The plot is depicted in Figure \ref{HC-IA_Hydrolysis}.

The conductivity curve can be divided into three regimes, labeled I, II, and III, each fitted with a linear function (blue lines). The carboxyl groups within the microgels are deprotonated until the crossover from regime I to regime II, as titration started at basic pH. In regime II they become progressively protonated. At the second crossover, from regime II to regime III, protonation of the microgels is complete. The concentration of negative charges within the microgels can be calculated from the volume of HCl consumed in regime II. From the points of intersection of the respective linear fits, a volume of 1.59~mL is derived to fully titrate the microgels, amounting to a concentration of negative charges of 3.10~mmol~g\textsuperscript{-1} (feed:~4.16~mmol~g\textsuperscript{-1}). Approximately 75\% of ester groups are transformed into pH-responsive acidic functionalities. Although the conversion is not quantitative, the microgel is still considered to be highly charged.

\subsection*{Assessment of the purification process}

\begin{figure}[H]
    \centering
    \includegraphics[scale=0.35]{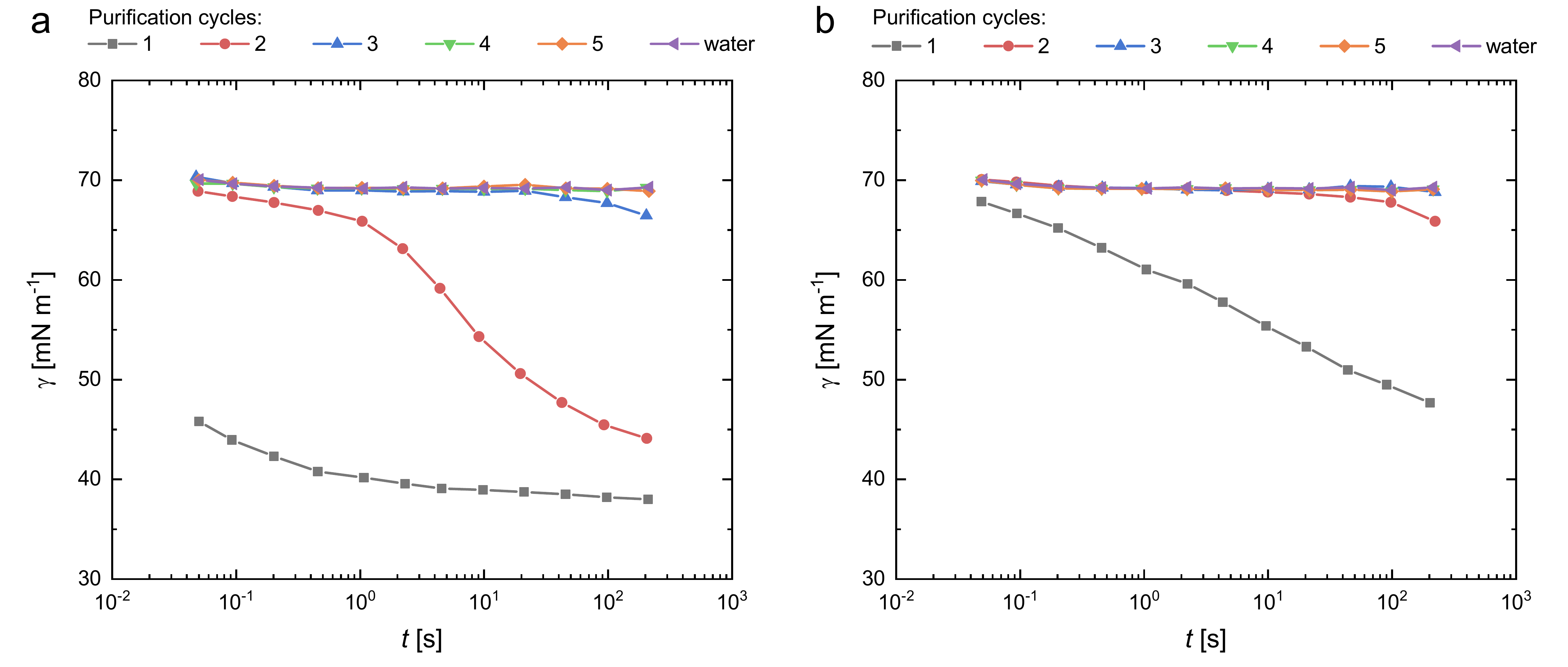}
    \caption{(a) Surface activity, surface tension $\gamma$ \textit{versus} surface age $t$, of the supernatant taken after each purification cycle after the initial synthesis of the uncharged precursor microgels. (b) Surface activity, surface tension $\gamma$ \textit{versus} surface age $t$, of the supernatant taken after each purification cycle after basic ester hydrolysis of the precursor microgels. Filtered ultra-pure water was measured as a reference.}
    \label{HC-IA_Purification}
\end{figure}

To investigate microgels in experiments involving fluid interfaces, they have to be free of surface-active impurities. Those impurities are likely remnants from the synthesis, e.g., monomers and oligomers or surfactants, that have not been removed during the purification procedure.\textsuperscript{46} We carefully assessed the quality of the purification process after the synthesis of the highly charged microgels. Therefore, we determined the surface tension $\gamma$ of the supernatant taken after each purification cycle with a bubble pressure tensiometer (BP100, KRÜSS GmbH Germany). All measurements were performed at room temperature, and samples of the supernatant were analyzed without further dilution or filtration. The results are included in Figure \ref{HC-IA_Purification}.

In both plots, the decrease of the surface tension with time becomes less with increasing number of purification steps, indicating the successive removal of interfacial active species. It remains constant over time after four cycles at a value of $\gamma$~$\approx$~70~mN~m\textsuperscript{-1} and is indistinguishable from the data of filtered ultra-pure water. A surface tension value of 70~mN~m\textsuperscript{-1} is close to the literature value determined for a clean interface between air and water at comparable temperatures of $\approx$~72~mN~m\textsuperscript{-1}.\textsuperscript{47} Slight deviations may be explained by small differences in temperature or the presence of dust particles, but they do not indicate the presence of interfacial-active substances. We conclude that our microgel solution is free of other interfacial-active substances. More generally, four centrifugation-redispersion cycles are sufficient to remove any impurities. Although only the HC-IA microgels were synthesized within the scope of this study, the other S-MAA and L-MAA microgels employed were purified in similar fashion and, therefore, are also considered free of contaminants.

\section*{S-MAA microgels at oil-water interfaces}

\subsection*{AFM height images}

\begin{figure}[H]
    \centering
    \includegraphics[scale=0.30]{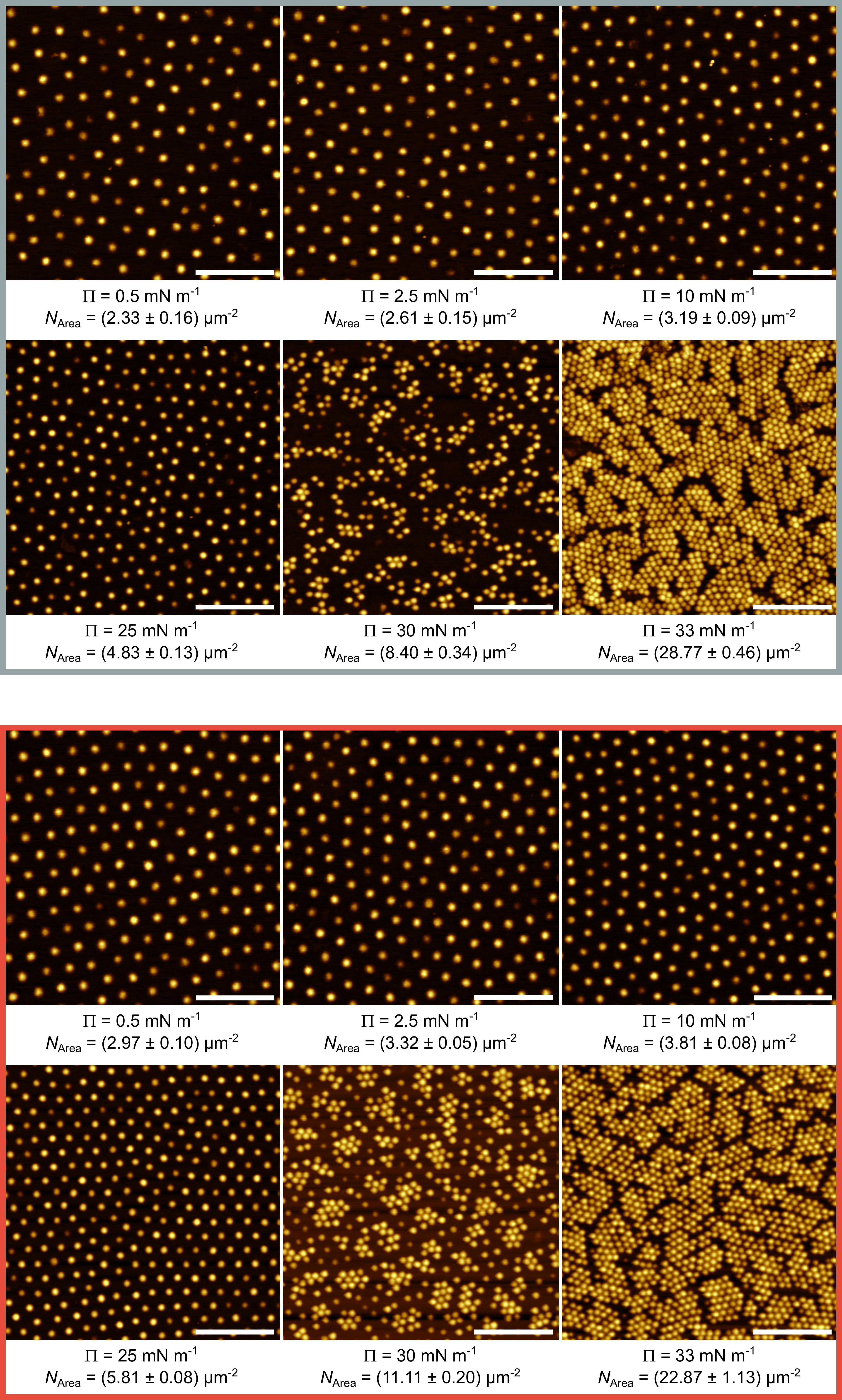}
    \caption{AFM height images of dried monolayers of the S-MAA microgels in the uncharged (grey box) and charged state (red box) deposited at controlled surface pressures $\Pi$. Scale bars are equal to 2~$\upmu$m. $N_{\text{Area}}$ is the number of microgels per area derived from quantitative image analysis.}
    \label{S-MAA_Images}
\end{figure}

\subsection*{Image analysis results}

\begin{table}[H]
    \footnotesize
    \centering
        \begin{threeparttable}
            \begin{tabular}{ccccccc}
            \toprule[1pt]
            & Regime & $\Pi$\textsuperscript{a} & $N_{\text{Area}}$\textsuperscript{b} & $d_{\text{cc,1}}$\textsuperscript{c} & $d_{\text{cc,2}}$\textsuperscript{c} & $\Psi_6$\textsuperscript{c}\\
            & & {[mN m\textsuperscript{-1}]} & {[$\upmu$m\textsuperscript{-2}]} & {[nm]} & {[nm]} &\\
            \midrule[1pt]
            \multirowcell{6}{uncharged \\ (pH 3, 0.1 mM KCl)} & II & 0.5 $\pm$ 0.3 & 2.33 $\pm$ 0.16 & 656 $\pm$ 72 & - & 0.50 $\pm$ 0.21\\
            & II & 2.5 $\pm$ 0.3 & 2.61 $\pm$ 0.15 & 618 $\pm$ 46 & - & 0.57 $\pm$ 0.22\\
            & II & 10 $\pm$ 0.3 & 3.19 $\pm$ 0.09 & 558 $\pm$ 38 & - & 0.56 $\pm$ 0.22\\
            & II & 25 $\pm$ 0.3 & 4.83 $\pm$ 0.13 & 446 $\pm$ 35 & - & 0.58 $\pm$ 0.22\\
            & III & 30 $\pm$ 0.3 & 8.40 $\pm$ 0.34 & 378 $\pm$ 100 & 195 $\pm$ 25 & 0.34 $\pm$ 0.15\\
            & IV & 33 $\pm$ 0.3 & 28.77 $\pm$ 0.46 & - & 163 $\pm$ 20 & 0.59 $\pm$ 0.24\\
            \addlinespace[5mm]
            \multirowcell{6}{charged \\ (pH 9, 0.1 mM KCl)} & II & 0.5 $\pm$ 0.3 & 2.97 $\pm$ 0.10 & 597 $\pm$ 22 & - & 0.93 $\pm$ 0.06\\
            & II & 2.5 $\pm$ 0.3 & 3.32 $\pm$ 0.05 & 571 $\pm$ 23 & - & 0.94 $\pm$ 0.06\\
            & II & 10 $\pm$ 0.3 & 3.81 $\pm$ 0.08 & 529 $\pm$ 24 & - & 0.93 $\pm$ 0.06\\
            & II & 10 $\pm$ 0.3 & 3.87 $\pm$ 0.13 & 528 $\pm$ 25 & - & 0.94 $\pm$ 0.05\\
            & II & 25 $\pm$ 0.3 & 5.81 $\pm$ 0.08 & 428 $\pm$ 20 & - & 0.93 $\pm$ 0.06\\
            & III & 30 $\pm$ 0.3 & 11.11 $\pm$ 0.20 & 387 $\pm$ 63 & 190 $\pm$ 23 & 0.43 $\pm$ 0.18\\
            & IV & 33 $\pm$ 0.3 & 22.87 $\pm$ 1.13 & - & 186 $\pm$ 18 & 0.71 $\pm$ 0.20\\
            \bottomrule[1pt]
            \end{tabular}
            \begin{tablenotes}
                \item \textsuperscript{a} Error of the instrument (film balance) was determined in a previous work and corresponds to the standard deviation from at least five independent measurements.\textsuperscript{40}
                \item \textsuperscript{b} Value is the arithmetic average obtained from multiple AFM images taken at different positions on the same substrate. Error corresponds to the standard deviation.
                \item \textsuperscript{c} Value is derived from Gaussian fit; Data from multiple AFM images taken at different positions on the same substrate are combined before fitting. Error corresponds to half the peak width of the Gaussian fitting function.
            \end{tablenotes}
        \end{threeparttable}
\caption{Results obtained from quantitative image analysis of dried monolayers of the S-MAA microgels deposited at controlled surface pressures $\Pi$. Data include the number of microgels per area $N_{\text{Area}}$, the center-to-center distance between microgels in the first or second hexagonal phase $d_{\text{cc,1/2}}$ and the hexagonal order parameter $\Psi_6$.}
\label{S-MAA_Data}
\end{table}

\subsection*{AFM phase images and diameter probability functions}

\begin{figure}[H]
    \centering
    \includegraphics[scale=0.35]{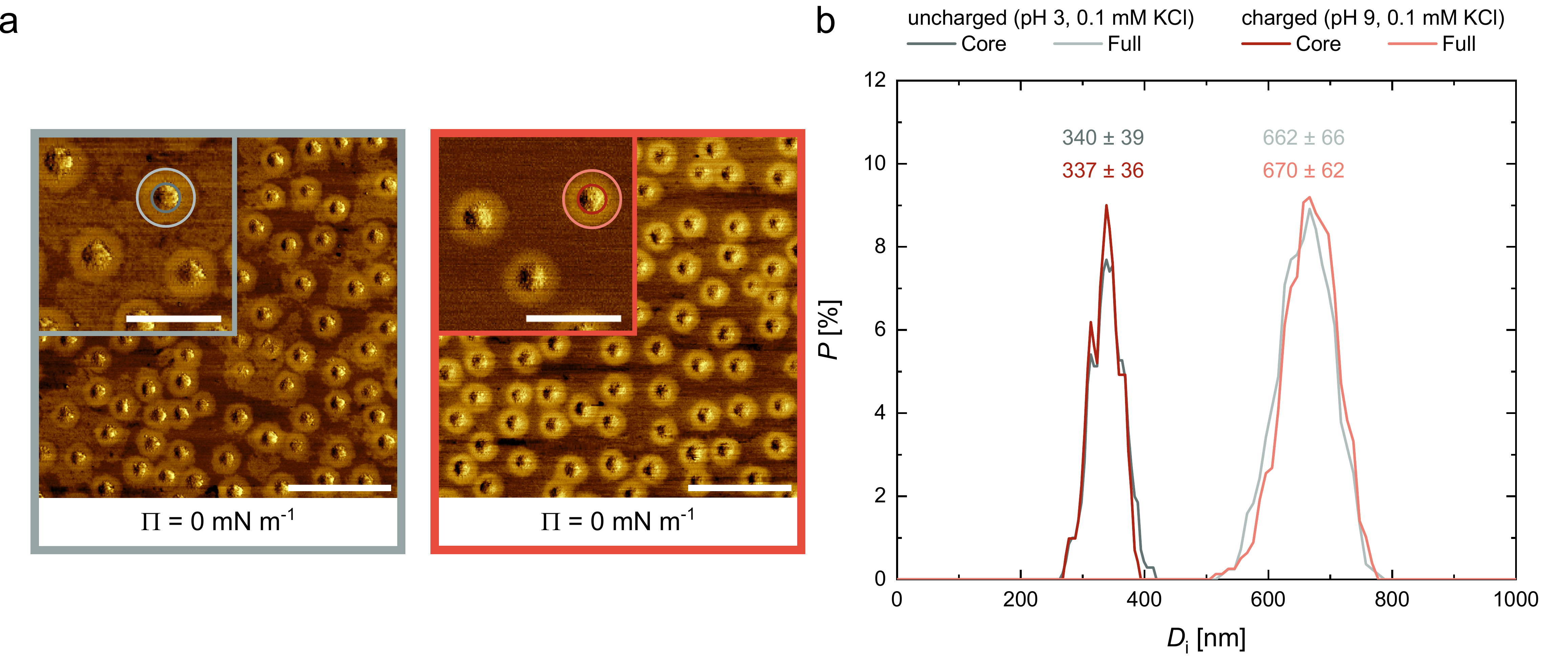}
    \caption{(a) AFM phase images of dried S-MAA microgels in the uncharged (grey box) and charged state (red box) deposited in dilute state at a surface pressure of $\Pi$~=~0~mN~m\textsuperscript{-1}. Insets are close-ups of individual microgels. Scale bars are equal to 2~$\upmu$m in the images and 1~$\upmu$m in the insets. (b) Diameter probability functions, probability $P$ \textit{versus} in-plane diameter $D_{\text{i}}$, of the S-MAA microgels in the uncharged and charged state. For both charge densities, the size of at least 100 individual microgels was determined from the images by hand and the data fitted with a Gaussian function (fits not shown).}
    \label{S-MAA_Images_Size2D}
\end{figure}

\subsection*{Short-range ordering}

\begin{figure}[H]
    \centering
    \includegraphics[scale=0.35]{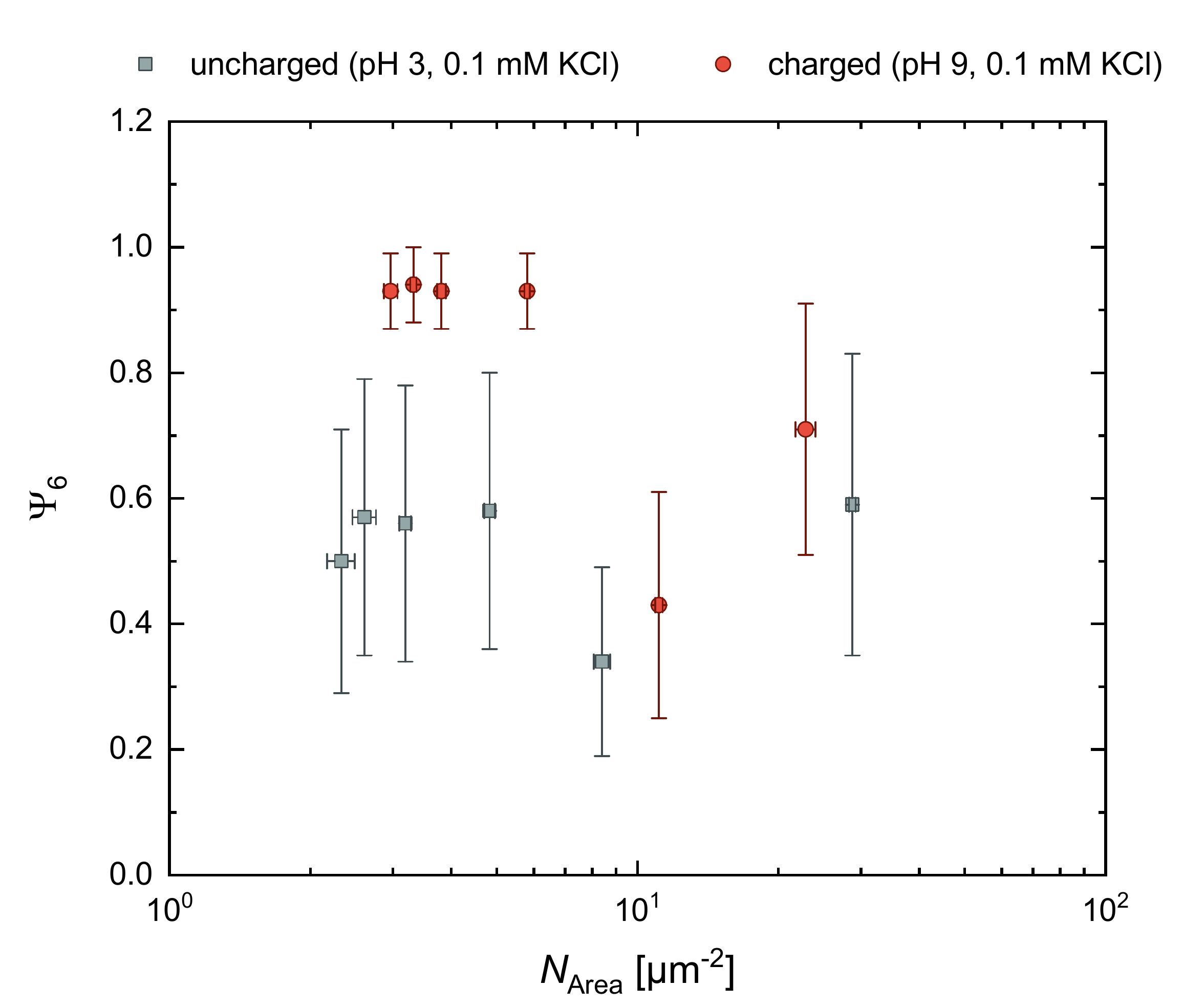}
    \caption{Short-range ordering, hexagonal order parameter $\Psi_6$ \textit{versus} number of microgels per area $N_{\text{Area}}$ in dried monolayers of the S-MAA microgels in the uncharged and charged state deposited at controlled surface pressures.}
    \label{S-MAA_Hexagonal_Order_Parameter}
\end{figure}

\clearpage

\section*{L-MAA microgels at oil-water interfaces}

\subsection*{AFM height images}

\begin{figure}[H]
    \centering
    \includegraphics[scale=0.30]{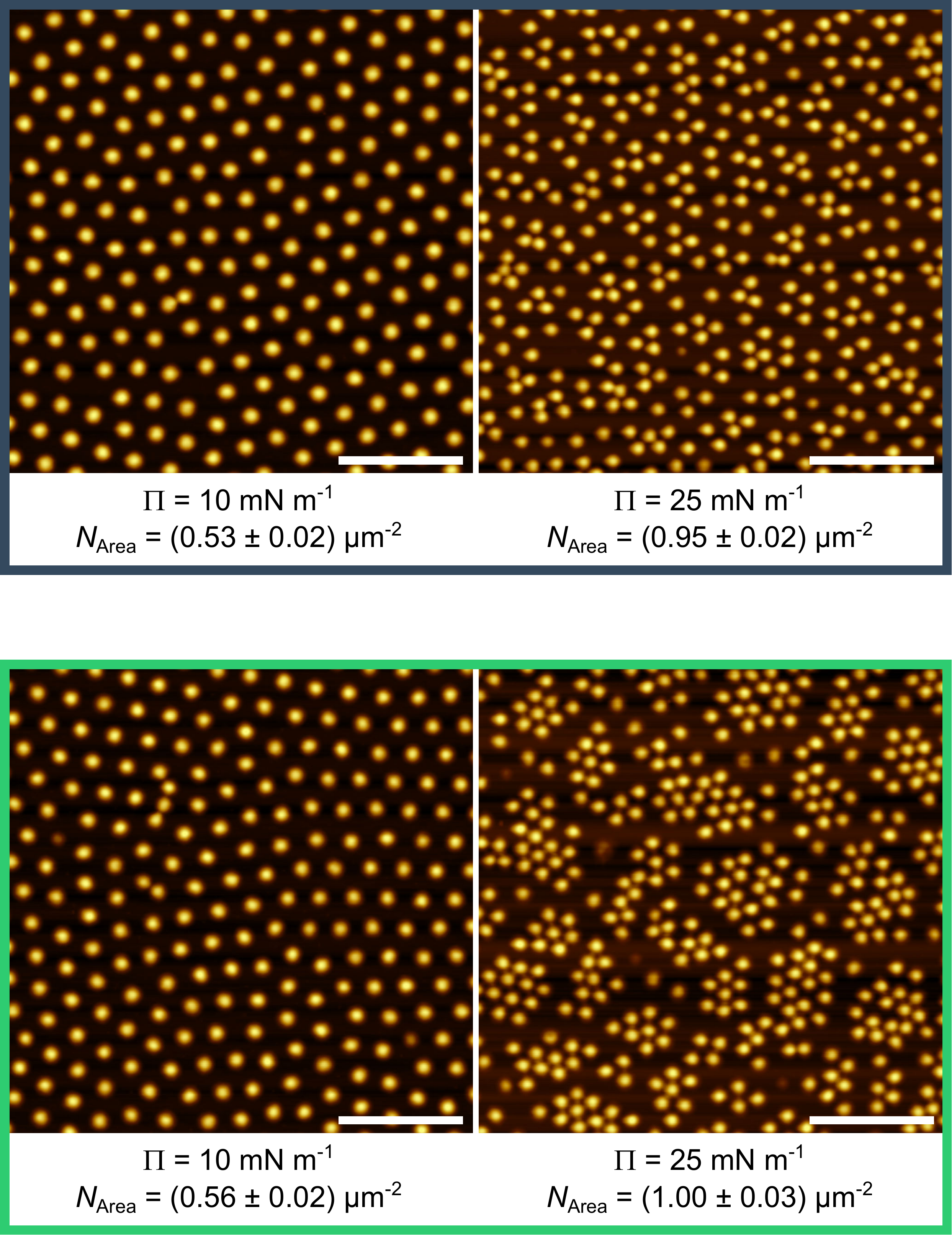}
    \caption{AFM height images of dried monolayers of the L-MAA microgels in the uncharged (dark grey box) and charged state (green box) deposited at controlled surface pressures $\Pi$. Scale bars are equal to 5~$\upmu$m. $N_{\text{Area}}$ is the number of microgels per area derived from quantitative image analysis.}
    \label{L-MAA_Images}
\end{figure}

\subsection*{Image analysis results}

\begin{table}[H]
    \footnotesize
    \centering
        \begin{threeparttable}
            \begin{tabular}{ccccccc}
            \toprule[1pt]
            & Regime & $\Pi$\textsuperscript{a} & $N_{\text{Area}}$\textsuperscript{b} & $d_{\text{cc,1}}$\textsuperscript{c} & $d_{\text{cc,2}}$\textsuperscript{c} & $\Psi_6$\textsuperscript{c}\\
            & & {[mN m\textsuperscript{-1}]} & {[$\upmu$m\textsuperscript{-2}]} & {[nm]} & {[nm]} &\\
            \midrule[1pt]
            \multirowcell{2}{uncharged \\ (pH 3, 0.1 mM KCl)} & II & 10 $\pm$ 0.3 & 0.53 $\pm$ 0.02 & 1412 $\pm$ 38 & - & 0.83 $\pm$ 0.23\\
             & III & 25 $\pm$ 0.3 & 0.95 $\pm$ 0.02 & 1204 $\pm$ 132 & 768 $\pm$ 66 & 0.45 $\pm$ 0.17\\
            \addlinespace[5mm]
            \multirowcell{2}{charged \\ (pH 9, 0.1 mM KCl)} & II & 10 $\pm$ 0.3 & 0.56 $\pm$ 0.02 & 1406 $\pm$ 44 & - & 0.96 $\pm$ 0.05\\
             & III & 25 $\pm$ 0.3 & 1.00 $\pm$ 0.03 & 1437 $\pm$ 167 & 817 $\pm$ 73 & 0.60 $\pm$ 0.22\\
            \bottomrule[1pt]
            \end{tabular}
            \begin{tablenotes}
                \item \textsuperscript{a} Error of the instrument (film balance) was determined in a previous work and corresponds to the standard deviation from at least five independent measurements.\textsuperscript{40}
                \item \textsuperscript{b} Value is the arithmetic average obtained from multiple AFM images taken at different positions on the same substrate. Error corresponds to the standard deviation.
                \item \textsuperscript{c} Value is derived from Gaussian fit; Data from multiple AFM images taken at different positions on the same substrate are combined before fitting. Error corresponds to half the peak width of the Gaussian fitting function.
            \end{tablenotes}
        \end{threeparttable}
\caption{Results obtained from quantitative image analysis of dried monolayers of the L-MAA microgels deposited at controlled surface pressures $\Pi$. Data include the number of microgels per area $N_{\text{Area}}$, the center-to-center distance between microgels in the first or second hexagonal phase $d_{\text{cc,1/2}}$ and the hexagonal order parameter $\Psi_6$.}
\label{L-MAA_Data}
\end{table}

\subsection*{Radial distribution functions}

\begin{figure}[H]
    \centering
    \includegraphics[scale=0.35]{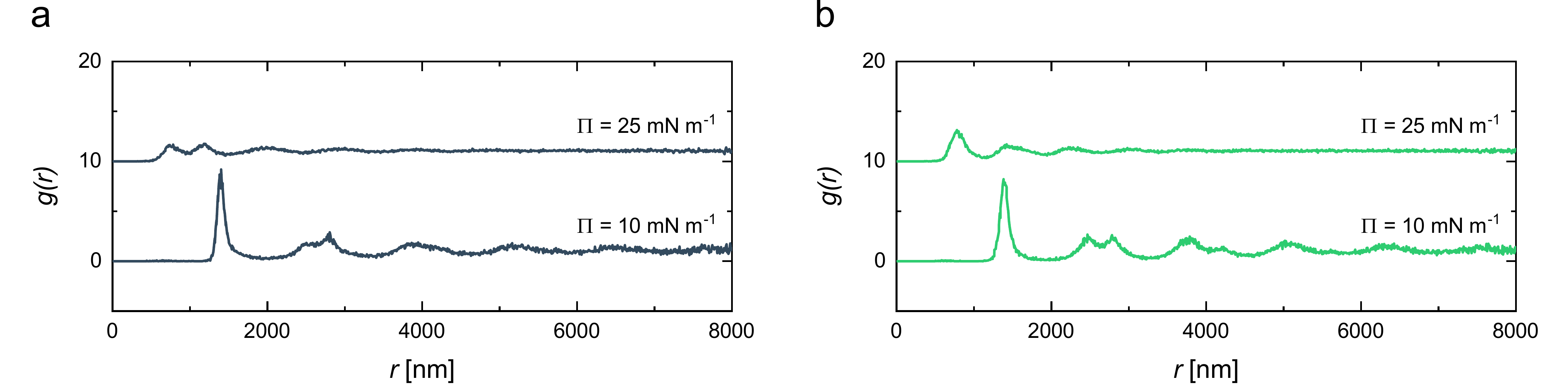}
    \caption{Radial distribution functions, \textit{g(r)} \textit{versus} distance \textit{r}, in dried monolayers of the L-MAA microgels in the (a) uncharged and (b) charged state deposited at controlled surface pressures $\Pi$.}
    \label{L-MAA_Radial_Distribution_Functions}
\end{figure}

\clearpage

\section*{HC-IA microgels at oil-water interfaces}

\subsection*{AFM height images}

\begin{figure}[H]
    \centering
    \includegraphics[scale=0.30]{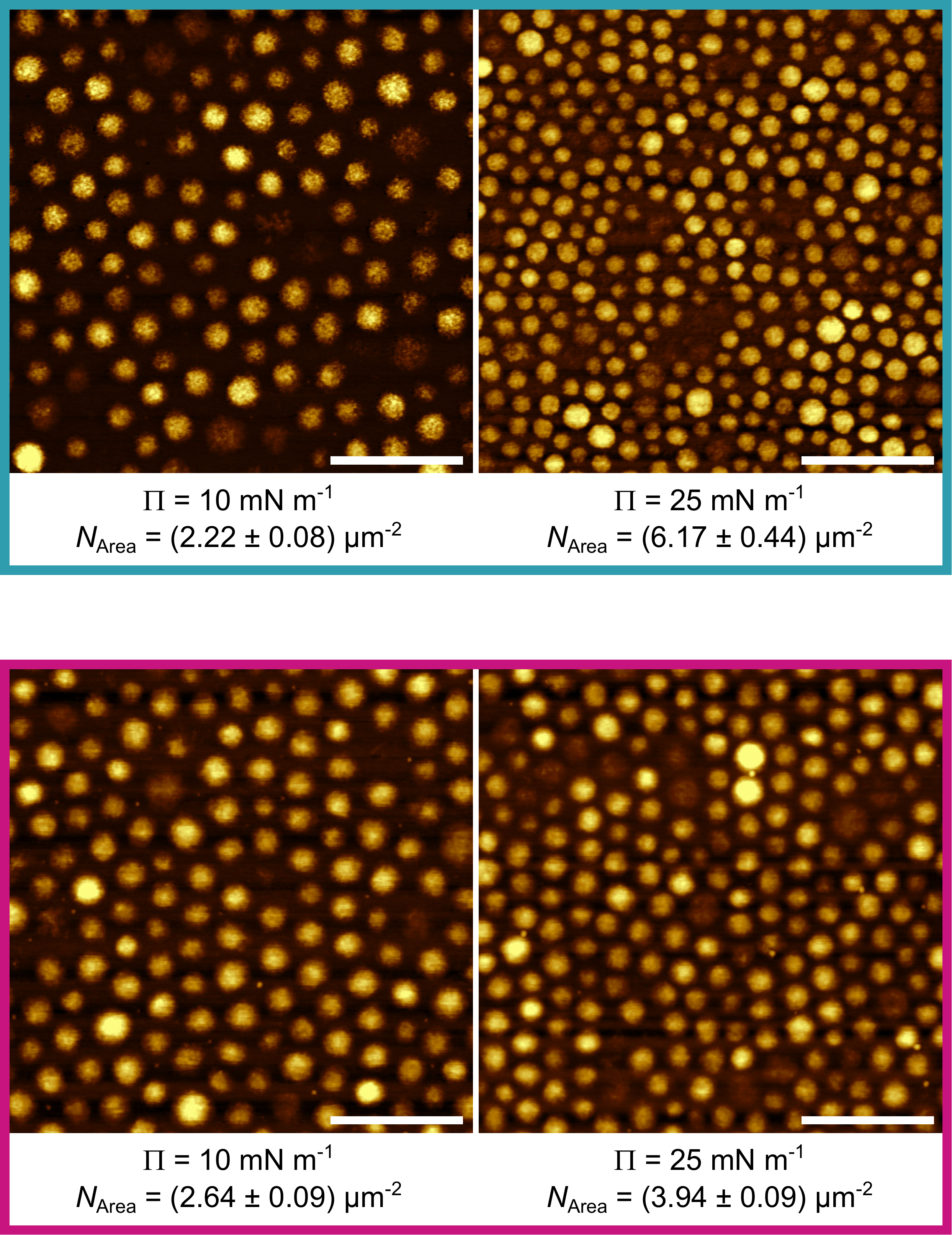}
    \caption{AFM height images of dried monolayers of the HC-IA microgels in the charged state at 100~mM KCl (blue box) and 0.1~mM KCl (violet box) deposited at controlled surface pressures $\Pi$. Scale bars are equal to 2~$\upmu$m. $N_{\text{Area}}$ is the number of microgels per area derived from quantitative image analysis.}
    \label{HC-IA_Images}
\end{figure}

\subsection*{Image analysis results}

\begin{table}[H]
    \footnotesize
    \centering
        \begin{threeparttable}
            \begin{tabular}{cccccc}
            \toprule[1pt]
            & $\Pi$\textsuperscript{a} & $N_{\text{Area}}$\textsuperscript{b} & $d_{\text{cc,1}}$\textsuperscript{c} & $d_{\text{cc,2}}$\textsuperscript{c} & $\Psi_6$\textsuperscript{c}\\
            & {[mN m\textsuperscript{-1}]} & {[$\upmu$m\textsuperscript{-2}]} & {[nm]} & {[nm]} &\\
            \midrule[1pt]
            \multirowcell{2}{charged \\ (pH 9, 100 mM KCl)} & 10 $\pm$ 0.3 & 2.22 $\pm$ 0.08 & 687 $\pm$ 70 & - & 0.50 $\pm$ 0.20\\
            & 25 $\pm$ 0.3 & 6.17 $\pm$ 0.44 & 423 $\pm$ 50 & - & 0.49 $\pm$ 0.21\\
            \addlinespace[5mm]
            \multirowcell{2}{charged \\ (pH 9, 0.1 mM KCl)} & 10 $\pm$ 0.3 & 2.64 $\pm$ 0.09 & 646 $\pm$ 55 & - & 0.65 $\pm$ 0.26\\
            & 25 $\pm$ 0.3 & 3.94 $\pm$ 0.09 & 534 $\pm$ 51 & - & 0.67 $\pm$ 0.23\\
            \bottomrule[1pt]
            \end{tabular}
            \begin{tablenotes}
                \item \textsuperscript{a} Error of the instrument (film balance) was determined in a previous work and corresponds to the standard deviation from at least five independent measurements.\textsuperscript{40}
                \item \textsuperscript{b} Value is the arithmetic average obtained from multiple AFM images taken at different positions on the same substrate. Error corresponds to the standard deviation.
                \item \textsuperscript{c} Value is derived from Gaussian fit; Data from multiple AFM images taken at different positions on the same substrate are combined before fitting. Error corresponds to half the peak width of the Gaussian fitting function.
            \end{tablenotes}
        \end{threeparttable}
\caption{Results obtained from quantitative image analysis of dried monolayers of the HC-IA microgels deposited at controlled surface pressures $\Pi$. Data include the number of microgels per area $N_{\text{Area}}$, the center-to-center distance between microgels in the first or second hexagonal phase $d_{\text{cc,1/2}}$ and the hexagonal order parameter $\Psi_6$.}
\label{HC-IA_Data}
\end{table}

\subsection*{Radial distribution functions}

\begin{figure}[H]
    \centering
    \includegraphics[scale=0.35]{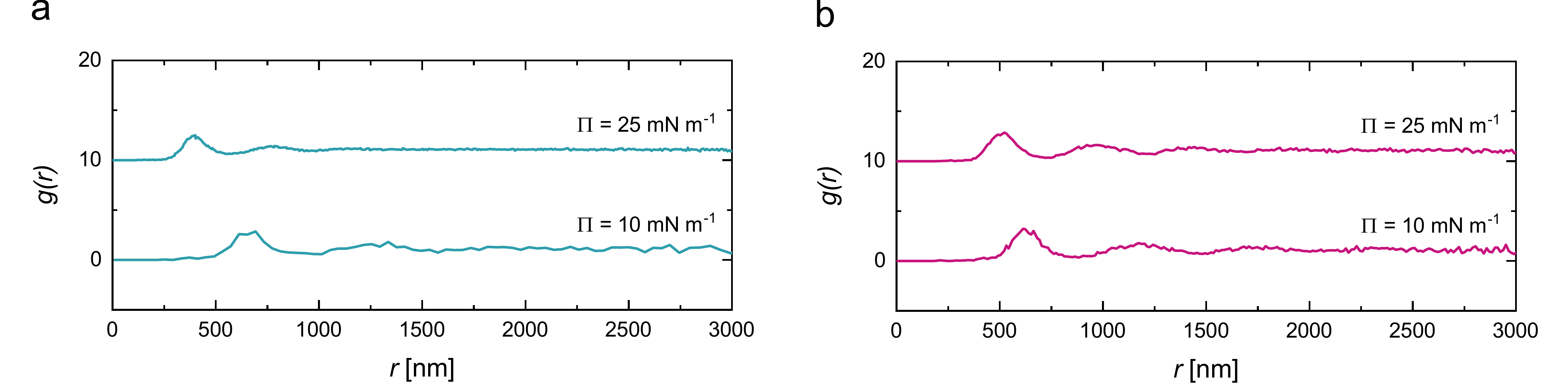}
    \caption{Radial distribution functions, \textit{g(r)} \textit{versus} distance \textit{r}, in dried monolayers of the HC-IA microgels in the charged state at (a) 100~mM KCl and (b) 0.1~mM KCl deposited at controlled surface pressures $\Pi$.}
    \label{HC-IA_Radial_Distribution_Functions}
\end{figure}

\clearpage

\section*{Computer simulations of microgels at the liquid-liquid interface}

\subsection*{Dependence of the effective microgel area on the interactions parameter}

\begin{figure}
    \centering
    \includegraphics[scale=0.35]{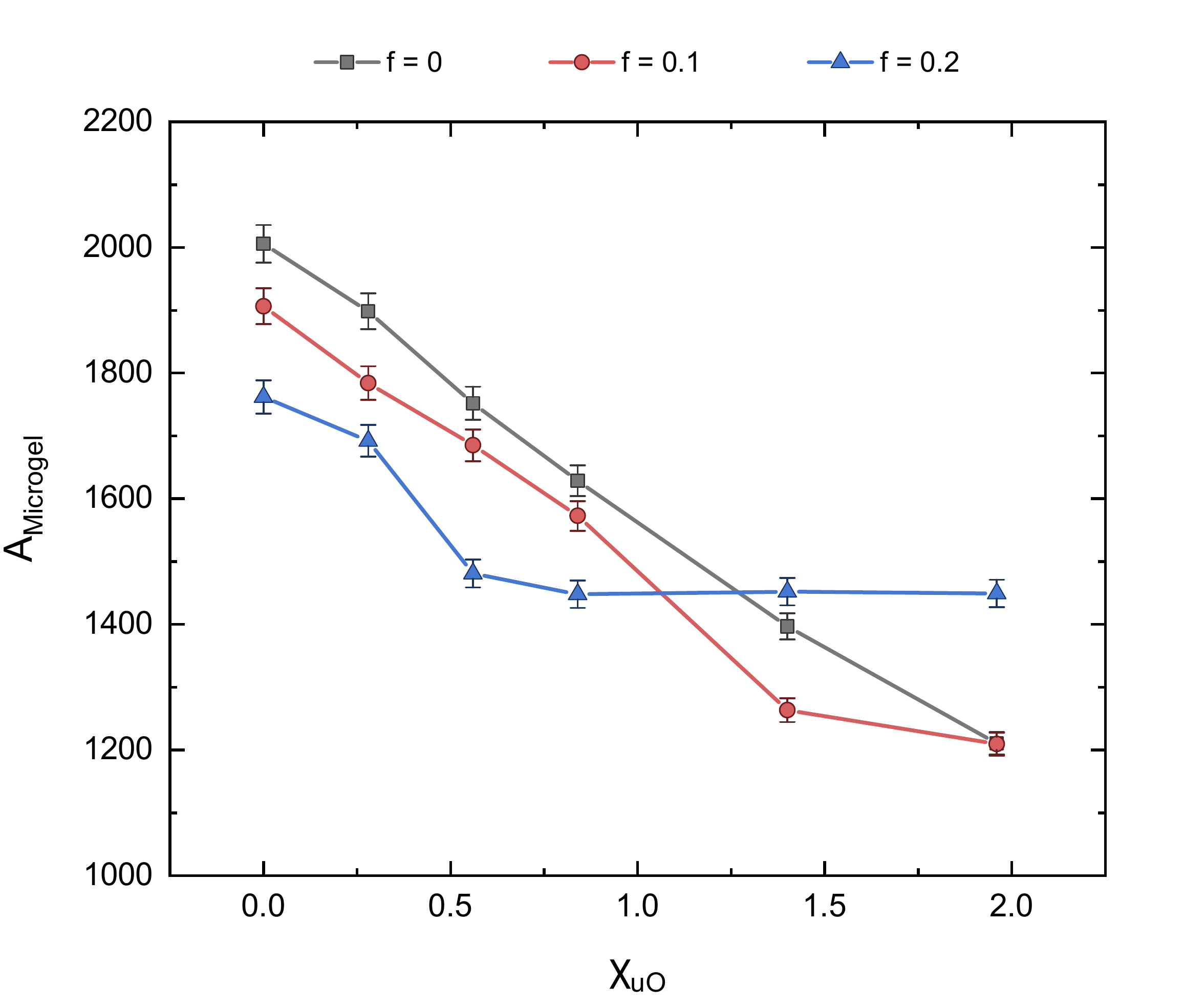}
    \caption{Dependence of the effective microgel area $A_{\text{Microgel}}$ (averaged over 3000 system configurations) on the $\chi_{\text{uO}}$ value for different fractions of charged monomer units $f$. The plateau observed for $f$~=~0.2 at $\chi_{\text{uO}}$~$\approx$~0.75 and for $f$~=~0.1 at $\chi_{\text{uO}}$~$\approx$~2 is related to the microgels being completely detached from the interface.}
    \label{SIM_Area}
\end{figure}